\input harvmac
\input psfig
\newcount\figno
\figno=0
\def\fig#1#2#3{
\par\begingroup\parindent=0pt\leftskip=1cm\rightskip=1cm\parindent=0pt
\global\advance\figno by 1
\midinsert
\epsfxsize=#3
\centerline{\epsfbox{#2}}
\vskip 12pt
{\bf Fig. \the\figno:} #1\par
\endinsert\endgroup\par
}
\def\figlabel#1{\xdef#1{\the\figno}}
\def\encadremath#1{\vbox{\hrule\hbox{\vrule\kern8pt\vbox{\kern8pt
\hbox{$\displaystyle #1$}\kern8pt}
\kern8pt\vrule}\hrule}}
\def\N{{\cal N}}
\overfullrule=0pt

%
\def\tilde{\widetilde}
\def\bar{\overline}
\def\Z{{\bf Z}}

\def\S{{\bf S}}
\def\R{{\bf R}}

\font\zfont = cmss10 

\def\bigone{\hbox{1\kern -.23em {\rm l}}}
\def\ZZ{\hbox{\zfont Z\kern-.4emZ}}

\Title{hep-th/9703166, IASSNS-HEP-97-19}
{\vbox{\centerline{SOLUTIONS OF FOUR-DIMENSIONAL FIELD }
\bigskip
\centerline{THEORIES VIA $M$ THEORY}}}
\smallskip
\centerline{Edward Witten\foot{Research supported in part
by NSF  Grant  PHY-9513835.}}
\smallskip
\centerline{\it School of Natural Sciences, Institute for Advanced Study}
\centerline{\it Olden Lane, Princeton, NJ 08540, USA}\bigskip

\medskip

\noindent
$\N=2$ supersymmetric gauge theories in four dimensions are studied by
formulating them as the quantum field theories derived from configurations of
fourbranes, fivebranes, and sixbranes in Type IIA superstrings, and then
reinterpreting those configurations in $M$ theory.  This approach leads to
explicit solutions for the Coulomb branch of a large family of 
four-dimensional $\N=2$ field theories with zero or negative beta function. 
\Date{March 1997}

\def\C{{\bf C}}

\newsec{Introduction}

Many interesting results about field theory and string theory have
been obtained by studying the quantum field theories that appear
 on the world-volume of string theory and $M$ theory branes.
One particular  construction that was considered recently
in $2+1$ dimensions \ref\hw{A. Hanany and E. Witten, ``Type IIB Superstrings,
BPS Monopoles, And Three-Dimensional Gauge Dynamics,''
 hep-th/9611230.} and has been further explored in  
\ref\oret{ Jan de Boer, Kentaro Hori, Hirosi Ooguri, Yaron Oz, Zheng Yin,
``Mirror Symmetry in 
Three-Dimensional Gauge Theories, $SL(2,Z)$ and D-Brane Moduli Spaces,'' 
hep-th/9612131,
Jan de Boer, Kentaro Hori, Yaron Oz, Zheng Yin, ``Branes And Mirror
Symmetry In $N=2$ Supersymmetric Gauge Theories In Three Dimensions,''
hep-th/9702154.}
and applied to
$\N=1$ models in four dimensions in
\ref\ketal{
S. Elitzur, A. Giveon, and D. Kutasov, ``Branes And $N=1$ Duality In String
Theory,'' hep-th/9702014.}
will be used
in the present paper to understand the Coulomb branch of some $\N=2$ models
in four dimensions.  The aim is to obtain for a wide class of four-dimensional
gauge theories with $\N=2$ supersymmetry the sort of description obtained
in \ref\swtwo{N. Seiberg and E. Witten, ``Electric-Magnetic Duality, Monopole
Condensation, And Confinement In $N=2$ Supersymmetric Yang-Mills Theory,''
Nucl. Phys. {\bf B426} (1994) 19,
``Monopoles, Duality, And Chiral Symmetry Breaking In $N=2$ Supersymmetric
QCD,'' Nucl. Phys. {\bf B341} (1994) 484.} for models with $SU(2)$ gauge group.

\midinsert
\centerline{\psfig{figure=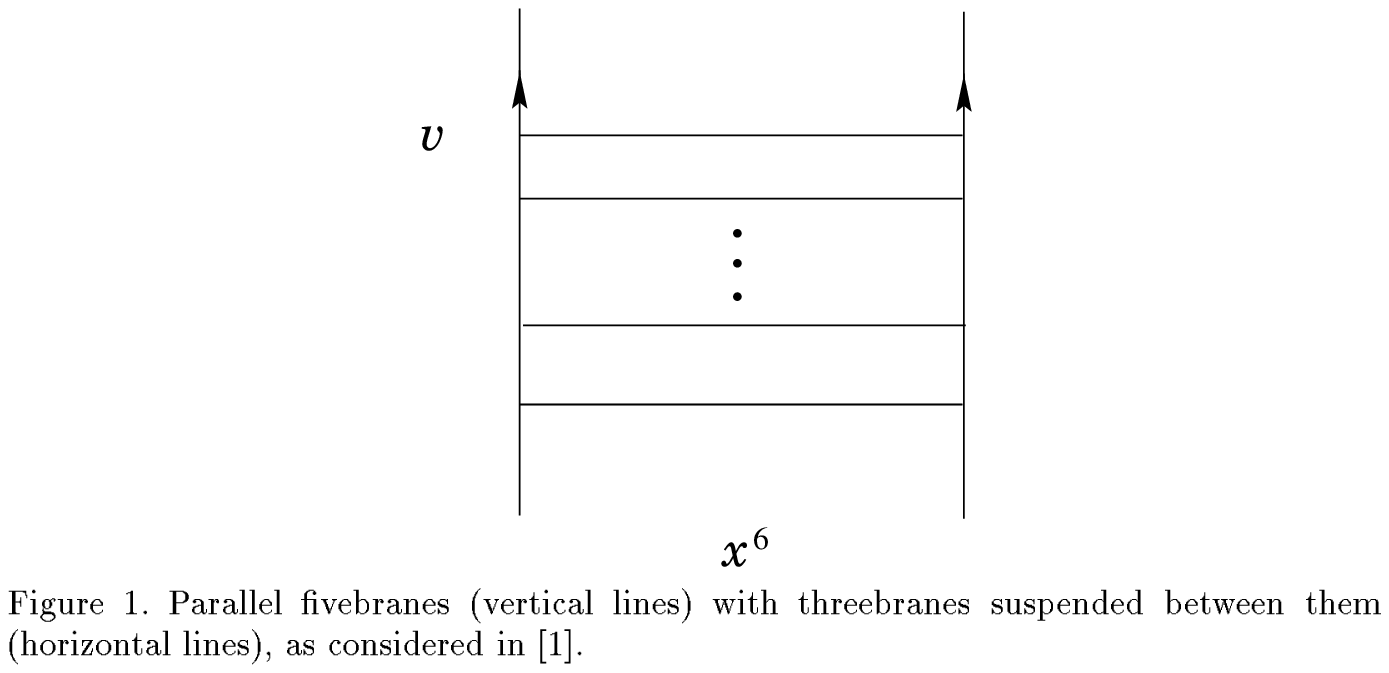,width=4.5in}}
\bigskip\bigskip
\endinsert

The construction in \hw\ involved branes of Type IIB superstring theory -- to 
be more
precise
the Dirichlet threebranes and the solitonic and Dirichlet fivebranes.
One considers, for example, 
NS fivebranes with threebranes suspended between them (figure 1).  The 
fivebranes,
being infinite in all six of their world-volume directions, are considered
to be very heavy and are treated classically.  The interest focusses on the
quantum field theory on the world-volume of the threebranes.
Being finite in one of their four dimensions, the threebranes are 
macroscopically
$2+1$ dimensional.  
The quantum field theory on this effective $2+1$ dimensional world
has eight conserved supercharges, corresponding to
$N=4$ supersymmetry in three dimensions or ${\cal N}=2$ in four
dimensions.  Many  properties of such a model can be effectively determined
using the description via branes.

To make a somewhat similar analysis of $3+1$ dimensional theories, 
one must replace the threebranes by fourbranes,
suspended between fivebranes (and, as it turns out, also in the presence of
sixbranes).  Since the fourbrane is infinite
in four dimensions (and finite in the fifth),  the field theory on such a 
fourbrane is
$3+1$-dimensional macroscopically.  

\def\Z{{\bf Z}}
\def\N{{\cal N}}
Type IIB superstring theory has no fourbranes, so we will consider Type IIA
instead.  Type IIA superstring theory has Dirichlet fourbranes, solitonic
fivebranes, and Dirichlet sixbranes.  Because there is only one brane of each
dimension, it will hopefully cause no confusion if we frequently drop the
adjectives ``Dirichlet'' and ``solitonic''   and
refer to the branes merely as fourbranes, fivebranes, and sixbranes.

One of the main techniques  in \hw\ was to use $SL(2,\Z)$ duality 
of Type IIB superstrings to predict a mirror symmetry of the $2+1$ dimensional
models.  For Type IIA there is no $SL(2,\Z)$ self-duality.  The strong
coupling limit of Type IIA superstrings in ten dimensions is instead determined
by an equivalence to eleven-dimensional $M$ theory; this equivalence
will be used in the present paper to obtain solutions of four-dimensional
field theories.  As we will see, a number of facts
about $M$ theory fit together neatly to make this possible.

\def\R{{\bf R}}
\def\S{{\bf S}}
In section 2, we explain the basic techniques and solve models that are 
constructed
from configurations of Type IIA fourbranes and fivebranes on ${\bf R}^{10}$.
In section 3, we incorporate sixbranes.
In section 4, we analyze models obtained by considering Type IIA
fourbranes and fivebranes
on ${\bf R}^9\times \S^1$.
\foot{Compactification of such a brane system on a circle has been considered
in \oret\ in the Type IIB context.}
  Many novel features will arise, including a 
geometric
interpretation of the gauge theory beta function in section 2 and a natural
family of conformally invariant theories in section 4. 
As we will see, each new step involves some essential new subtleties, though
formally the brane diagrams are analogous (and related by $T$-duality) to those
 in \hw.

\newsec{Models With Fourbranes And Fivebranes}

In this section we consider fourbranes suspended between fivebranes in
Type IIA superstring theory on ${\bf R}^{10}$.  Our fivebranes will
be located at $x^7=x^8=x^9=0$ and -- in the classical approximation --
at some fixed values of $x^6$.  
The worldvolume of the fivebrane is parametrized by the values of the remaining
coordinates $x^0,x^1,\dots,x^5$.

\midinsert
\centerline{\psfig{figure=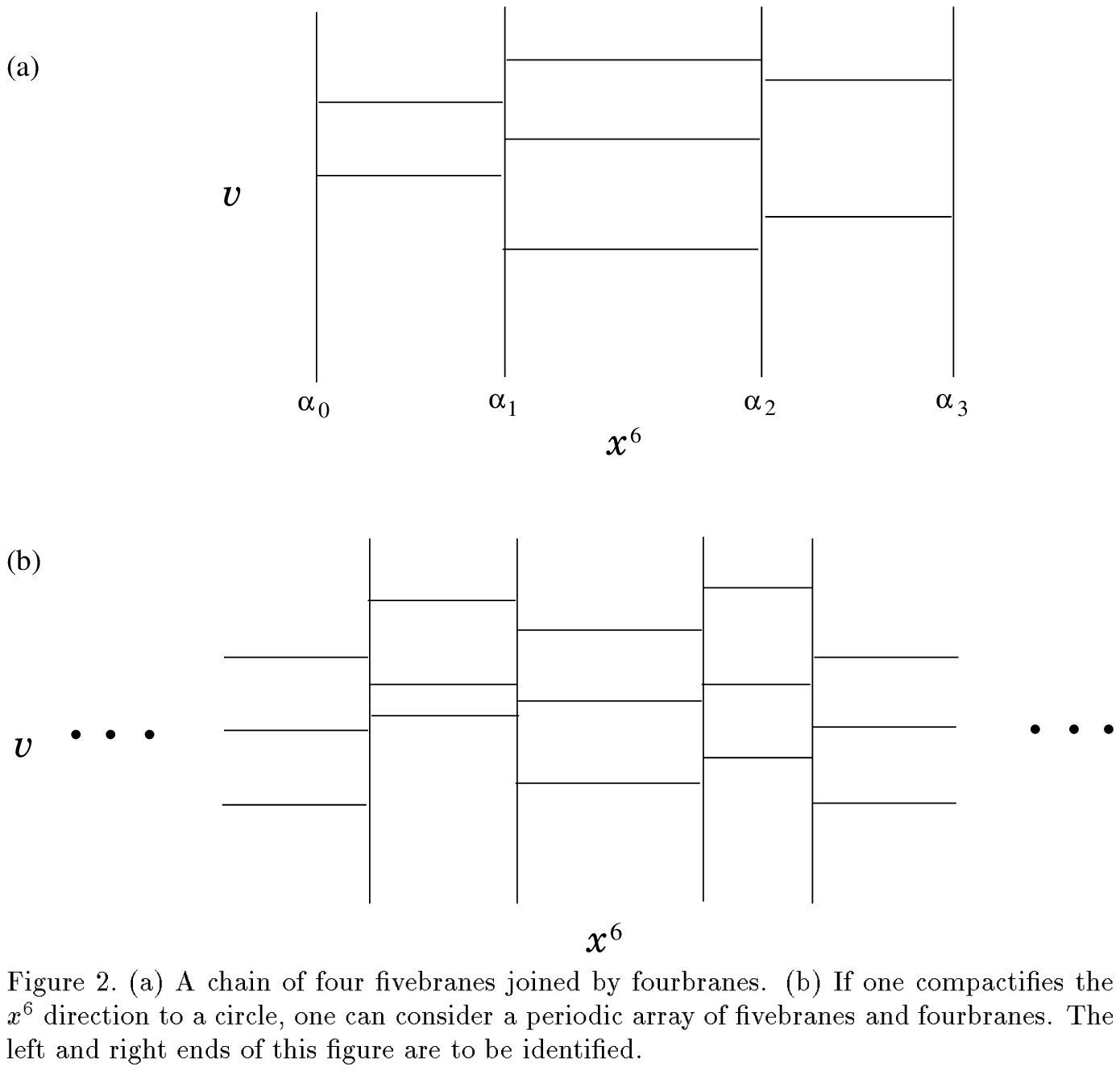,width=3.5in}}
\bigskip\bigskip\bigskip
\endinsert
In addition, we introduce fourbranes whose world-volumes are parametrized
by $x^0,x^1,x^2,x^3$, and $x^6$.  However, our fourbranes will not be infinite
in the $x^6$ direction.  They will terminate on fivebranes. (Occasionally
we will consider a semi-infinite fourbrane that terminates on a fivebrane at
one end, and extends to $x^6=\infty$ or $-\infty$ at the other end.)  A typical
picture is thus that of figure 2(a).  As in \hw, we will examine this picture first from
the fivebrane point of view and then from the fourbrane point of view.

It will be convenient to introduce a complex variable $v=x^4+ix^5$.   
Classically,
every fourbrane is located at a definite value of $v$.
The same is therefore also true for its possible endpoints on a fivebrane.

\subsec{Theory On Fivebrane}

A fact that was important in \hw\ is that on the worldvolume of
a Type IIB fivebrane there propagates a $U(1)$ gauge field.  A system of
$k$ parallel but noncoincident
fivebranes can be interpreted as a system with $U(k)$ gauge
symmetry spontaneously broken to $U(1)^k$.  Points at which Type IIB
threebranes end on fivebranes carry magnetic charge in this spontaneously
broken gauge theory.

Even though one draws similar brane pictures in the Type IIA case, the
interpretation is rather different.  Type IIA fivebranes do not carry gauge
fields, but rather self-dual antisymmetric tensors.  When parallel fivebranes
become coincident, one gets not enhanced gauge symmetry but a strange
critical point with tensionless strings 
\ref\strom{
A. Strominger, ``Open $p$-Branes,'' 
hep-th/9512059.}, 
concerning
which too little is known for it to be useful in the present paper.

However, the endpoints of a fourbrane on a fivebrane do behave as charges
in an appropriate sense.  A fivebrane on which fourbranes 
end does not really have a definite value of $x^6$ as the classical brane
picture suggests.  The fourbrane ending on a fivebrane creates a ``dimple'' in
the fivebrane.  What one would like to call the $x^6$ value of the fivebrane
is really the $x^6$ value measured at $v=\infty$, far from the disturbances
created by the fourbranes.

To see whether this makes sense, note that $x^6$ is determined as a function 
of $v$
by minimizing the total fivebrane worldvolume.  For large $v$ the equation
for $x^6$ reduces to a Laplace equation,
\eqn\togo{\nabla^2 x^6=0.}
Here $\nabla^2$ is the Laplacian on the fivebrane worldvolume.  $x^6$ is a 
function
only of the directions normal to the fourbrane ends, that is only of $v$ and 
$\bar v$.
Since the Green's function of the Laplacian in two dimensions is a logarithm,
the large $v$ behavior of $x^6$ is determined by \togo\ to be
\eqn\bogo{x^6=k\ln |v| +{\rm constant}}
for some $k$.  Thus, in general, there is no well-defined large $v$ limit of
$x^6$.  This contrasts with the situation considered in \hw\ where (because
of considering threebranes instead of fourbranes) $x^6$ obeys a {\it 
three}-dimensional
Laplace equation, whose solution approaches a constant at infinity.  The 
limiting
value $x^6(\infty)$ is then the ``$x^6$ value of the fivebrane''
which appears in the classical brane diagram and was used in \hw\ to 
parametrize
the configurations.

Going back to the Type IIA case, for a fivebrane with a single fourbrane ending
on it from, say, the left, 
$k$ in \bogo\ is an absolute constant that depends only  on the fourbrane
and fivebrane tensions (and hence the Type IIA string coupling constant).
However, a fourbrane ending on a fivebrane on its right pulls in the opposite
direction and contributes to $k$ with the opposite sign from a fivebrane 
ending on
the left.   If $a_i$, $i=1,\dots, q_L$ and $b_j$, $j=1,\dots, q_R$ are the
$v$ values of fourbranes that end on a given fivebrane on its left and
on its right, respectively, then the asymptotic form of $x^6$ is
\eqn\jogo{x^6=k\sum_{i=1}^{q_L}\ln |v-a_i| -k\sum_{j=1}^{q_R}\ln |v-b_j|+
{\rm constant}.    }
We see that $x^6$ has a well-defined limiting value for $v\to \infty$ if
and only if $q_L=q_R$, that is if there are equal forces on the fivebrane from
both left and right.

For any finite chain of fivebranes with fourbranes ending on them,
as in figure 2(a), it is impossible to obey this condition, assuming that
there are no semi-infinite fourbranes that go off to $x^6=\infty$ or
$x^6=-\infty$.  At least the fivebranes at the ends of the chain are subject
to unbalanced forces.  The ``balanced'' case, a chain of fivebranes each 
connected
by the same number of fourbranes, as in figure 2(b), 
is most natural if one compactifies the $x^6$
direction to a circle, so that all fourbranes are finite in extent.  
It is very special and will be the subject of section 4.

Another important question  is affected by a related infrared divergence.
For this, we consider the motion of fourbranes.
When the fourbranes move, the disturbances they produce on the fivebranes
move also, producing a contribution to the fourbrane kinetic energy.   We
consider a situation in which the $a_i$ and $b_j$ vary as a function of
the first four coordinates $x^\mu$, $\mu=0,\dots ,3$ (which are the 
``spacetime''
coordinates of the effective four-dimensional field theories studied in this 
paper).
The fivebrane kinetic energy has a term $\int d^4x d^2v 
\sum_{\mu=0}^3\partial_\mu
x^6\partial^\mu x^6$.
With $x^6$ as in \jogo, this becomes
\eqn\judo{k^2\int d^4x\,d^2v \left|{\rm Re}\left(\sum_i\partial_\mu a_i \left({1\over
v-a_i}\right)- \sum_j\partial_\mu b_j \left({1\over
v-b_j}\right)\right)\right|^2.}
The $v$ integral converges if and only if
\eqn\boggo{\partial_\mu\left(\sum_ia_i-\sum_jb_j\right)=0,}
so that
\eqn\koggo{\sum_ia_i-\sum_jb_j=q_\alpha,}
where $q_\alpha$ is a constant characteristic of the
$\alpha^{th}$  fivebrane.  
While the $q_\alpha$ are constants that we will
eventually interpret in terms of  ``bare masses,'' the remaining $a$'s and 
$b$'s
are free to vary; they are indeed ``order parameters'' which depend on the
choice of quantum vacuum of the four-dimensional field theory.

\def\N{{\cal N}}
The above discussion of the large $v$ behavior of $x^6$ and its kinetic energy
is actually only half of the story.  From the point of view of the  
four-dimensional
$\N=2$ supersymmetry of our brane configurations, $x^{6}$ 
is the real part of a complex field that is in a vector multiplet.  
The imaginary part of this superfield is a scalar field that propagates on
the fivebrane.  If Type IIA superstring theory on $\R^{10}$ is reinterpreted
as $M$ theory on $\R^{10}\times \S^1$, the scalar in question is the position
of the fivebrane in the eleventh dimension.  We have labeled the
ten dimensions of Type IIA as $x^0,x^1,\dots,x^9$, so we will call the eleventh
dimenson $x^{10}$.
In generalizing \jogo\ to include $x^{10}$, we will use  $M$ theory units 
(which differ by a Weyl rescaling from Type IIA units used in \jogo).  Also,
we understand $x^{10}$ to be a periodic variable with period $2\pi R$.

With this understood, the generalization of \jogo\ to include $x^{10}$ is
\eqn\illop{x^6+ix^{10} =R
\sum_{i=1}^{q_L}\ln (v-a_i) -R\sum_{j=1}^{q_R}\ln (v-b_j)+
{\rm constant}.  }
The fact that $x^6+ix^{10}$ varies holomorphically with $v$ is required by 
supersymmetry.  The imaginary part of this equation 
states that $x^{10}$ jumps by $\pm 2\pi R$ 
when one circles around one of the $a_i$ or $b_j$ in the complex $v$ plane.
In other words, the endpoints of fourbranes on a fivebrane behave as vortices
in the fivebrane effective theory (an overall  constant in \illop\ was fixed
by requiring that the vortex number is one).
This is analogous, and related by $T$-duality,
to the fact that the endpoint of a threebrane on a fivebrane looks like a
magnetic monopole, with magnetic charge one,  in the fivebrane theory;
this  fact was extensively used in \hw.  
The interpretation of brane boundaries as charges on other branes was
originally described in \strom.

In terms of $s=(x^6+ix^{10})/R$, the last formula reads
\eqn\millop{s=\sum_{i=1}^{q_L}\ln (v-a_i)-\sum_{j=1}^{q_R}\ln(v-b_j)+{\rm 
constant}.}

\subsec{Four-Dimensional Interpretation}

Now we want to discuss what the physics
on this configuration of branes looks like to a four-dimensional observer.

We consider a situation, shown in figure 2(a) in a special case, 
with $n+1$ fivebranes, lebeled
by $\alpha=0,\dots, n$.  Also,
for $\alpha=1,\dots, n$, we include $k_\alpha$ fourbranes between the 
$(\alpha-1)^{th}$ 
and $\alpha^{th}$ fivebranes.  

It might seem that the gauge group would be $\prod_{\alpha=1}^n U(k_\alpha)$, 
with
each   $U(k_\alpha)$  factor coming from the corresponding set of $k_\alpha$ 
parallel
fourbranes.  However, \koggo\ means precisely that the $U(1)$ factors
are ``frozen out.''  To be more precise, in \koggo, $\sum_ia_i$ is the
scalar part of the $U(1)$ vector multiplet in one factor $U(k_{\alpha})$,
and $\sum_jb_j$ is the scalar part of the $U(1)$ multiplet in the ``next''
gauge group factor $U(k_{\alpha+1})$.
\koggo\ means that the difference $\sum_ia_i-\sum_jb_j$ is ``frozen,'' and
therefore, by supersymmetry, an entire $U(1)$ vector supermultiplet is actually
missing from the spectrum.  Since such freezing occurs at each point in the 
chain,
including the endpoints (the fivebranes with fourbranes ending only on one 
side),
the $U(1)$'s are all frozen out and the gauge group is actually 
$\prod_{\alpha=1}^n SU(k_\alpha)$.    

\def\k{{\bf k}}
What is the hypermultiplet spectrum in this theory?  By reasoning exactly as in
\hw, massless hypermultiplets arise (in the classical approximation of the 
brane
diagram) precisely when fourbranes end on a fivebrane from opposite sides
at the same point in spacetime.  Such a hypermultiplet is charged precisely 
under
the gauge group factors coming from fourbranes that adjoin the given fivebrane.
So the hypermultiplets transform, in an obvious notation, as
$(\k_1,\bar \k_2) \oplus (\k_2,\bar\k_3) \oplus \dots\oplus 
(\k_{n-1},\bar\k_n)$.
The constants $q_\alpha$ in \koggo\ determine the bare masses $m_\alpha$ of 
the 
$(\k_\alpha,\bar \k_{\alpha+1})$ hypermultiplets, so in fact arbitrary bare
masses are possible.  The bare masses are actually
\eqn\jumper{m_\alpha={1\over k_\alpha}\sum_i
a_{i,\alpha}-{1\over k_{\alpha+1}}\sum_ja_{j,\alpha+1},}
where $a_{i,\alpha}$, $i=1,\dots , k_\alpha$  are the positions in the  $v$ 
plane
of the fourbranes between the $\alpha-1^{th}$ and $\alpha^{th}$ fivebrane. 
In other words, $m_\alpha$ is the difference between the average position in
the $v$ plane of the fourbranes to the left and right of the $\alpha^{th}$ fivebrane.
$m_\alpha$ is not simply a multiple of $q_\alpha$, but the $q_\alpha$ for
$\alpha=1,\dots,n$ determine the $m_\alpha$.

Now, we come to the question of what is the coupling constant of the 
$SU(k_\alpha)$
gauge group.  Naively, if $x^6_\alpha$ is the $x^6$ value of the $\alpha^{th}$ 
fivebrane, then the gauge coupling $g_\alpha$ of $SU(k_\alpha)$ should be given
by 
\eqn\ibbo{{1\over g_\alpha^2} = {x^6_\alpha-x^6_{\alpha-1} \over \lambda},}
where $\lambda$ is the string coupling constant.

We have here a problem, though.  What precisely is meant by the objects 
$x^6_\alpha$?
As we have seen above, these must be understood as functions of $v$ which in 
general
diverge for $v\to\infty$.  Therefore, we must interpret $g_\alpha$ as a 
function
of $v$:
\eqn\nibbo{{1\over g_\alpha^2(v) } ={x^6_\alpha(v)-x^6_{\alpha-1}(v) 
\over\lambda}.}
We interpret $v$ as  setting a mass scale, and $g_\alpha(v)$ as the effective
coupling of  the $SU(k_\alpha)$ theory at mass $|v|$.  Then $1/g_\alpha^2(v)$ 
generally, according to \jogo, diverges logarithmically for $v\to\infty$.
But that is familiar in four-dimensional gauge theories: the inverse gauge 
coupling
of an asymptotically free theory diverges logarithmically at high energies.
We thus interpret this divergence as reflecting the one loop beta function
of the four-dimensional theory.  

It is natural to include $x^{10}$ along with $x^6$, and thereby to get a 
formula for the effective theta angle $\theta_\alpha$ of the $SU(k_\alpha)$
gauge theory, which is determined by the separation in the $x^{10}$ direction
between the $\alpha-1^{th}$ and $\alpha^{th}$ fivebranes. Set 
\eqn\omibo{\tau_\alpha={\theta_\alpha\over 2\pi}+{4\pi i\over g_\alpha^2}.}
Then in terms of $s=(x^6+ix^{10})/R$ (with distances now
measured in $M$ theory units) we have 
\eqn\tomibo{-i\tau_\alpha(v) =s_\alpha(v) - s_{\alpha-1}(v).} 
(A multiplicative constant on the right hand side has been set to one by
requiring that under $x^{10}_\alpha\to x^{10}_\alpha+2\pi R$, the theta
angle changes by $\pm 2\pi$.)
But according to  \millop,   at large $v$ one has $s_\alpha(v)=(
k_{\alpha}-k_{\alpha+1})\ln v$, so 
\eqn\omibo{-i\tau_\alpha(v) \cong (2k_\alpha-k_{\alpha-1}-k_{\alpha+1})\ln v.}

The standard asymptotic freedom formula is $-i\tau = b_0\ln v$, where
$-b_0$ is the coefficient of the one-loop beta function.  So \omibo\ amounts
to the statement that the one-loop beta function for the $SU(k_\alpha)$ factor
of the gauge group is
\eqn\pogo{b_{0,\alpha}=-2k_\alpha+k_{\alpha-1}+k_{\alpha+1}.}
This is in agreement with a standard field theory computation for this model.
  In fact, for $\N=2$ supersymmetric QCD with gauge group
$SU(N_c)$ and $N_f$ flavors, one usually has $b_0=-(2N_c-N_f)$.  In the
case at hand, $N_c=k_\alpha$, and the $(\k_{\alpha-1},\bar\k_\alpha)\oplus
(\k_\alpha,\bar \k_{\alpha+1})$ hypermultiplets
make the same contribution to the $SU(k_\alpha)$
beta function as $k_{\alpha-1}+k_{\alpha+1}$ flavors, so the effective value
of $N_f$ is $k_{\alpha-1}+k_{\alpha+1}$.

\subsec{Interpretation Via $M$ Theory} 

By now we have identified a certain class of models that can be constructed
with fivebranes and fourbranes only.  The remaining question is of course
how to analyze these models.  For this we will use $M$ theory.

First of all, the reason that one may effectively go to $M$ theory is
that according to \ibbo, a rescaling of the Type IIA string coupling constant,
if accompanied by a rescaling of the separations of the fivebranes in the $x^6$
direction, does not affect the field theory coupling constant and so 
is irrelevant.  One might be concerned that \ibbo\ is just a classical
formula.
But in fact, we have identified in the brane diagram all marginal and relevant
operators (the coupling constants and hypermultiplet bare masses) of the low
energy $\N=2$ field theory, so any additional parameters 
(such as the string coupling constant) really are irrelevant.
Therefore we may go to the regime of large $\lambda$.

\def\S{{\bf S}}
What will make this  useful is really the following.
A fourbrane ending on a fivebrane has no known explicit conformal field theory
description. The end of the fourbrane is a kind of singularity that is hard
to understand in detail. That is part of the limitation of describing this 
system via
Type IIA superstrings.  But in $M$ theory everything we need can be explicitly 
understood
using only the low energy limit of the theory.
The Type IIA fivebrane on $\R^{10}$ is simply an $M$ theory fivebrane
on $\R^{10}\times \S^1$, whose world-volume, roughly, is located at a point in 
$\S^1$
and spans a six-manifold in $\R^{10}$.  A Type IIA fourbrane
is an $M$ theory fivebrane that is wrapped over the $\S^1$ (so that, roughly,
its world-volume projects to a five-manifold in $\R^{10}$).
Thus, {\it the four-brane and five-brane come from the same basic object
in $M$ theory}.  The Type IIA singularity where the fourbrane appears to end
on a fivebrane is, as we will see, completely eliminated by going to $M$ 
theory.

The Type IIA configuration of parallel fivebranes joined by fourbranes
can actually be reinterpreted in $M$ theory
as a configuration of a single fivebrane with a more complicated
world history. The fivebrane world-volume
 be described as follows.  (1) It
sweeps out arbitrary values of the first four coordinates $x^0,x^1,\dots, x^3$.
It is located at $x^7=x^8=x^9=0$.  (2)  In the remaining four coordinates
$x^4,x^5,x^6$, and $x^{10}$ -- which parametrize a four-manifold $Q\cong 
\R^3\times
\S^1$ -- the fivebrane worldvolume spans a two-dimensional
surface $\Sigma$.  (3) If one forgets $x^{10}$ and projects to a Type IIA
description in terms of branes on $\R^{10}$, then one gets back, in the limit
of small $R$, the classical configuration of fourbranes and fivebranes that
we started with.
(4) Finally,
$\N=2$  supersymmetry means that if we give $Q$ the complex structure in which
$v=x^4+ix^5$ and $s=x^6+ix^{10}$ are holomorphic, then $\Sigma$ is a complex
Riemann surface in $Q$.  This makes $\R^4\times \Sigma$ a supersymmetric
cycle in the sense of \ref\bekstrom{K. Becker, M. Becker, and A. Strominger,
``Fivebranes, Membranes, And Nonperturbative String Theory,''
    Nucl. Phys. {\bf B456} (1995) 130.} and so ensures spacetime supersymmetry.

In the approximation of the Type IIA brane diagrams,
$\Sigma$ has different components that are described locally by saying that
$s$ is constant (the fivebranes) or that $v$ is constant (the fourbranes).
But the singularity that appears in the Type IIA limit where the different
components meet can perfectly well
be absent upon going to $M$ theory; and that will be so generically, as we will
see.  Thus, for generic values of the parameters, $\Sigma$ will be
a smooth complex Riemann surface in $Q$.

This smoothness is finally the reason that going to $M$ theory leads to
a solution of the problem.  For large $\lambda$, all distances characteristic
of the Riemann surface $\Sigma$ are large and it will turn out that
there are generically no singularities.
So  obtaining and analyzing the solution
will require only a knowledge of the low energy long wavelength approximation to
$M$ theory and its fivebranes.

\bigskip\noindent{\it Low Energy Effective Action}

We will now work out the low energy four-dimensional physics that will
result from such an $M$ theory configuration.  The discussion is analogous to, 
but more
elementary than, a situation considered in \ref\vwmore{A. Klemm, W. Lerche,
P. Mayr, C. Vafa, and N. Warner, ``Self-Dual Strings And $N=2$ Supersymmetric
Field Theory,'' 
hep-th/9604034.}
where an $\N=2$ theory in four dimensions was related to a brane of the general
form $\R^4\times \Sigma$.

Vector multiplets will appear in four dimensions because on the worldvolume
of an $M$ theory fivebrane there is a chiral antisymmetric tensor field 
$\beta$,
that is, a two-form $\beta$ whose three-form field strength $T$ is self-dual.
Consider in general a fivebrane whose
 worldvolume is $\R^4\times \bar \Sigma$, where $\bar\Sigma$
is a compact Riemann surface of genus $g$.  According to 
\ref\verlinde{E. Verlinde, ``Global Aspects Of Electric-Magnetic Duality,''
hep-th/9506011.}, in
the effective four-dimensional description, 
the zero modes of the antisymmetric tensor give $g$ abelian gauge fields
on $\R^4$.  The coupling constants and theta parameters of the $g$ abelian
gauge fields are described by a rank $g$ abelian variety which is simply
the Jacobian $J(\bar \Sigma)$.  

These conclusions are reached as follows.  Let
\eqn\hogo{T=F\wedge \Lambda +*F\wedge *\Lambda,}
where $F$ is a two-form on $\R^4$, $\Lambda$ is a one-form on $\bar\Sigma$, and
$*$ is the Hodge star.  This $T$ is self-dual, and the equation of motion
$dT=0$ gives Maxwell's equations $dF=d*F=0$ along with the equations $d\Lambda
=d*\Lambda=0$ for $\Lambda$.  So $\Lambda$ is a harmonic one-form, and
every choice of a harmonic one-form $\Lambda$ gives a way of embedding
solutions of Maxwell's equations on $\R^4$ as solutions of the equations
for the self-dual three-form $T$.  If $\bar\Sigma$ has genus $g$, then
the space of self-dual (or anti-self-dual) $\Lambda$'s is $g$-dimensional,
giving $g$ positive helicity photon states (and $g$ of negative helicity)
on $\R^4$.  The low energy theory thus has gauge group $U(1)^g$.
The terms quadratic in $F$  in the
effective action for the gauge fields  are obtained by inserting \hogo\
in the fivebrane kinetic energy $\int_{\R^4\times \bar\Sigma}|T|^2$;
the Jacobian of $\bar\Sigma$ enters by determining the integrals of
wedge products of $\Lambda$'s and $*\Lambda$'s.

\def\CP{{\bf CP}}
In our problem of $n+1$ parallel Type IIA fivebranes joined by fourbranes,
the $M$ theory fivebrane is $\R^4\times \Sigma$, where $\Sigma$ is not 
compact.  So
the above discussion does not immediately apply.
However, $\Sigma$ could be compactified by adding $n+1$ points.  Indeed, for
a single fivebrane, $\Sigma$ would be a copy of $\C$ (the $v$ plane), which
is $\CP^1$ with a point deleted.   So if there are no fourbranes, we have
just $n+1$ disjoint copies of $\CP^1$ with a point omitted from each.
Including a fourbrane means cutting holes
out of adjoining fivebranes and connecting them with a tube.
This produces (if all $k_\alpha$ are positive) a connected Riemann surface
$\Sigma$ which can be compactified by adding $n+1$ points.  Note that the 
deleted
points are ``at infinity''; the metric on $\Sigma$ that is obtained from its
embedding in $Q$ is complete and looks ``near each puncture'' like the
flat complex plane with the puncture being the point  ``at infinity.''

The reason that noncompactness potentially modifies the discussion of
the low energy effective action is that in \hogo, one must ask for
$\Lambda$ to be square-integrable, in the metric on $\Sigma$ which comes
from its embedding in $Q$, as well as harmonic.  Since the punctures are ``at 
infinity,''
square-integrability implies that $\Lambda$ has vanishing periods on
a contour that surrounds any puncture.  (A harmonic one-form $\Lambda'$ 
that has a non-vanishing
period on such a contour would look near $v=\infty$ like $\Lambda'=dv/v$,
leading to $\int \Lambda'\wedge *\Lambda'=\int dv\wedge d\bar v/|v|^2=\infty$.)
Hence $\Lambda$ extends over the compactification $\bar\Sigma$ of $\Sigma$.
Since moreover the equation for a one-form on $\bar \Sigma$
to be self-dual is conformally invariant
and depends only on the complex structure of $\bar\Sigma$, the 
square-integrable
harmonic one-forms on $\Sigma$ are the {\it same} as the harmonic one-forms
on $\bar\Sigma$.  So finally, in our problem, the low energy effective
action of the vector fields is determined by the Jacobian of $\bar\Sigma$.

It is thus of some interest to determine the genus of $\bar\Sigma$.
We construct $\bar \Sigma$ beginning with $n+1$ disjoint copies of $\CP^1$,
of total Euler characteristic $2(n+1)$.  Then we glue in a total of
$\sum_{\alpha=1}^n k_\alpha$ tubes between adjacent $\CP^1$'s.  Each time
such a tube is glued in, the Euler characteristic is reduced by two, so the
final value is $2(n+1) -2\sum_{\alpha=1}^nk_\alpha$.  This equals $2-2g$,
where $g$ is the genus of $\bar\Sigma$.  So we get 
$g=\sum_{\alpha=1}^n(k_\alpha-1)$.
This is the expected dimension of the Coulomb branch for the gauge group
$\prod_{\alpha=1}^nSU(k_\alpha)$.  In particular, this confirms that
the $U(1)$'s are ``missing''; for the gauge group to be 
$\prod_{\alpha=1}^nU(k_\alpha)$,
the genus would have to be $\sum_{\alpha=1}^nk_\alpha$.  

So far we have emphasized the effective action for the four-dimensional gauge
fields.  Of course, the rest of the effective action is determined from this
via supersymmetry.  For instance, the scalars in the low energy effective
action simply determine the embedding of $\R^4\times \Sigma$ in spacetime,
or more succinctly 
the embedding of $\Sigma$ in $Q$; and their kinetic energy is obtained by
evaluating the kinetic energy for motion of the fivebrane in spacetime.

\bigskip\noindent{\it The Integrable System}

In general, the low energy effective action for an $\N=2$ system in four
dimensions is determined by an integrable Hamiltonian system in the complex 
sense.
The expectation values of the scalar fields in the vector multiplets
are the commuting Hamiltonians; the orbits generated by the commuting
Hamiltonian flows are the complex tori which determine the kinetic energy of
the massless four-dimensional vectors.  This structure was noticed in
special cases 
\nref\gorsky{A. Gorsky, I. Krichever, A. Marshakov, A. Mironov, and A. Morozov,
``Integrability and Seiberg-Witten Exact Solution,'' hep-th/9505035, Nucl. Phys.
{\bf B459} (1996) 97.}
\nref\mw{E. Martinec and N. Warner,  ``Integrable Systems And Supersymmetric
Gauge Theory,'' hep-th/9509161, Nucl. Phys. {\bf B459} (1996) 97.}
\nref\nakatsu{
T. Nakatsu and K. Takasaki, ``Whitham-Toda Hierarchy And $N=2$ Supersymmetric
Yang-Mills Theory,''
hep-th/9509162, Mod. Phys. Lett. {\bf A11} (1996) 157.}
\refs{\gorsky -
\nakatsu}
and deduced from the generalities
of low energy supersymmetric effective field theory \ref\donagiwitten{R. Donagi 
and
E. Witten, ``Supersymmetric Yang-Mills Theory And Integrable Systems,''
hep-th/9510101.}. 

 A construction of many complex integrable systems is as follows.
Let $X$ be a two-dimensional complex symplectic manifold.  Let $\Sigma$
be a complex curve in $X$.  Let ${\cal W}$ be the
 deformation space of pairs $(\Sigma',{\cal L}')$,
where $\Sigma'$ is a curve in $X$ to which $\Sigma$ can be deformed and ${\cal 
L}'$ is
a line bundle on $\Sigma'$   of specified degree.
Then ${\cal W}$ is  an integrable system; it has
a complex symplectic structure which is such that 
any functions that depend only on the choice of $\Sigma'$ (and not of ${\cal 
L}'$)
are Poisson-commuting.  The Hamiltonian flows generated by these 
Poisson-commuting
functions are the linear motions on the space of ${\cal L}'$'s, that is, on the
Jacobian of $\Sigma'$. 

This integrable system was described in 
\nref\odo{R. Donagi, L. Ein, and R. Lazarsfeld, ``A Non-Linear Deformation Of
The Hitchin Dynamical System,'' alg-geom/9504017.}
\odo, as a generalization of a gauge theory construction
by Hitchin \ref\hitchin{N. Hitchin, ``The Self-Duality Equations On
A Riemann Surface,'' Proc. London Math. Soc. {\bf 55} (1987) 59,
``Stable Bundles And
Integrable Systems,'' Duke Math. J. {\bf 54} (1987) 91.}; a prototype for
the case of non-compact $\Sigma$ is the extension of
Hitchin's construction to Riemann surfaces with punctures in
\ref\markman{E. Markman, ``Spectral
Curves And Integrable Systems,'' Comp. Math. {\bf 93} (1994) 255.}.   
The same integrable system
has appeared in the description of certain BPS states for Type IIA superstrings
on K3 \ref\vafabersa{M. Bershadsky, V. Sadov, and C. Vafa,
``$D$-Branes And Topological Field Theories,''
hep-th/9511222.}.

In general, fix a hyper-Kahler metric on the complex symplectic manifold $X$
(of complex dimension two)
and consider $M$ theory on $\R^7\times X$.  Consider a fivebrane of the
form $\R^4\times \Sigma$, where $\R^4$ is a fixed linear subspace of $\R^7$ 
(obtained
by setting three linear combinations of the seven coordinates to constants) and
$\Sigma$ is a complex curve in $X$.  Then the effective $\N=2$ theory on $\R^4$
is controlled by the integrable system described in the last paragraph, with 
the given
$X$ and $\Sigma$.  This follows from the fact that the scalar fields in the
four-dimensional theory parametrize the choice of a curve $\Sigma'$ to which $\Sigma$
can be deformed (preserving its behavior at infinity) while the  Jacobian of
$\Sigma'$ determines the couplings of the vector fields.

The case of immediate interest is the case that $X=Q$
and $\Sigma$ is related to the brane diagram with which we started the present
section.  The merit of this case (relative to an arbitrary pair $(X,\Sigma)$) 
is
that because of the Type IIA interpretation, we know a gauge theory
whose solution is given by this special case of the integrable model.
Some generalizations that involve different choices of $X$ are in sections
3 and 4.

\bigskip\noindent{\it BPS States}

\nref\town{P. Townsend,
``$D$-Branes From $M$-Branes,''
hep-th/9512062,  Phys.Lett.{\bf  B373} (1996) 68-75.}
The spectrum of massive BPS states in models constructed
this way can be analyzed roughly as in \vwmore, by
using the fact that $M$ theory twobranes can end on fivebranes 
\refs{\strom,\town}.
BPS states can be obtained by considering suitable twobranes in $\R^7\times X$.
To ensure the BPS property,
 the twobrane world volume should be a product $\R\times D$, where
$\R$ is a straight line in $\R^4\subset \R^7$ (representing ``the world line
of the massive particle in spacetime'') and $D\subset X$ is a complex Riemann 
surface with
a non-empty boundary $C$ that lies on $\Sigma$.  
By adjusting  $D$ to minimize the area of $D$ (keeping fixed the holomogy
class of $C\subset \Sigma$), one gets a twobrane worldvolume whose quantization
gives a BPS state.

\subsec{Solution Of The Models}

We now come to the real payoff, which is the solution of the models.

The models are to be described in terms of an equation $F(s,v)=0$, defining
a complex curve in $Q$.

Since $s$ is not single-valued, we introduce 
\eqn\introt{t=\exp(-s)=\exp(-(x^6+ix^{10})/R)} and look
for an equation $F(t,v)=0$.

Now if $F(t,v)$ is regarded as a function of $t$ for fixed $v$, then
the roots of $F$ are the positions of the fivebranes (at the given value of 
$v$).
The degree of $F$ as a polynomial in $t$ is therefore the number of fivebranes.
To begin with, we will consider a model with only two fivebranes.
$F$ will therefore be quadratic in $t$.

Classically, if one regards $F(t,v)$ as a function of $v$ for fixed $t$,
with a  value of $t$ that is ``in between'' the two fivebranes,
then the roots of $F(t,v)$ are the values of $v$ at which there are fourbranes.
We will set the number of fourbranes suspended between the two
fivebranes equal to $k$, so $F(t,v)$ should be
of degree $k$ in $v$.  (If $t$ is ``outside'' the classical position
of the fivebranes, the polynomial
$F(t,v)$ still vanishes for $k$ values of $v$; these roots will occur at large
$v$ and are related to the ``bending'' of the fivebranes for large $v$.) 

So such a model
will be governed by a curve of the form
\eqn\uvv{A(v)t^2+B(v)t+C(v)=0,}
with $A,B,$ and $C$ being polynomials in $v$ of degree $k$. We set 
$F=At^2+Bt+C$.

At a zero of $C(v)$, one of the roots of \uvv\ (regarded as an equation for 
$t$)
goes to $t =0$.  According to \introt, $t=0$ is $x^6=\infty$.  
Having a root of the equation which goes to $x^6=\infty$ at a fixed limiting
value of $v$ (where  $C(v)$ vanishes) means that there is a semi-infinite
fourbrane to the ``right'' of all of the fivebranes. 

Likewise, at a zero of $A(v)$, a root of  $F$ goes to $t=+\infty$, that
is to say, to $x^6=-\infty$.  This corresponds to a semi-infinite fourbrane
on the ``left.''  

Since there are $k$ fourbranes between the two fivebranes, these theories
will be $SU(k)$ gauge theories.
As in \hw, a semi-infinite fourbrane, because of its infinite extent
in $x^6$, has an infinite kinetic energy (relative to the fourbranes that 
extend
a finite distance in $x^6$) and can be considered to be frozen in place at a 
definite value of $v$.  The effect of a semi-infinite fourbrane is to add
one hypermultiplet in the fundamental representation of $SU(k)$.

\nref\argyres{P. Argyres and A. Faraggi,  ``The Vacuum Structure And Spectrum
Of $N=2$ Supersymmetric $SU(N)$ Gauge Theory,''
hep-th/9411057,  Phys. Rev. Lett. {\bf 74} (1995) 3931.}
\nref\more{
S. Klemm, W. Lerche, S. Theisen, and S. Yankielowicz,
``Simple Singularities and $N=2$ Supersymmetric Yang-Mills Theory,''
hep-th/9411048, Phys. Lett. {\bf B344} (1995) 169-175.}
We first explore the ``pure gauge theory'' without hypermultiplets.
For this we want no zeroes of $A$ or $C$, so $A$ and $C$ must be constants
and the curve becomes after a rescaling of $t$ 
\eqn\unvu{t^2+B(v)t+1=0.}
In terms of $\tilde t=t+B/2$, this reads
\eqn\joppo{\tilde t^2={B(v)^2\over 4}-1.}
By rescaling and shifting $v$, 
one can put $B$ in the form
\eqn\polip{B(v)=v^k+u_2v^{k-2}+u_3v^{k-3}+\dots+u_k.}
\joppo\ is our first success; it
is a standard form of the curve that governs the $SU(k)$ theory without
hypermultiplets \refs{\argyres,\more}.

We chose $F(t,v)$ to be of degree $k$ in $v$ so that, for a value of $t$ that
corresponds to being ``between'' the fivebranes, there would be $k$ roots
for $v$.  Clearly, however, the equation $F(t,v)=0$ has $k$ roots for $v$
for {\it any} non-zero $t$ (we recall that $t=0$ is ``at infinity'' in the
original variables).  What is the interpretation of these roots for
very large or very small $v$, to the left or right of the fivebranes?
For $t$ very large, the roots for $v$ are approximately at
\eqn\imop{t\cong c\cdot \,v^k,}
and for $t$ very small they are approximately at
\eqn\omopit{t\cong c'\cdot\,v^{-k};}
here  $c,c'$ are constants.
The formulas $t\cong v^{\pm k}$ are actually special cases of \millop; they
represent the ``bending'' of the fivebranes as a result of being pulled
on by fourbranes.  The formulas \imop\ and \omopit\ show that for 
$x^6\to \pm \infty$, the roots of $F$, as  a function of $v$ for fixed $t$,
 are at very large $|v|$.  These
roots do not correspond, intuitively, to positions of fourbranes
but are points ``near infinity'' on the bent fivebranes.

We can straightforwardly incorporate hypermultiplet flavors in this discussion.
For this, we merely incorporate zeroes of $A$ or $C$.
For example, to include $N_f$ flavors we can take $A=1$ and
$C(v) =f\prod_{j=1}^{N_f}(v-m_j)$ where the $m_j$, being the zeroes of
$C$, are the positions
of the semi-infinite fourbranes or in other words the hypermultiplet bare 
masses,
and $f$ is a complex constant.
Equation \joppo\ becomes 
\eqn\yuyu{\tilde t^2={B(v)^2\over 4}-f\prod_{j=1}^{N_f}(v-m_j).}
We set now
\eqn\guffo{B(v)=e(v^n+u_2v^{n-2}+u_3v^{n-3}+\dots + u_n)}
with $e$ and the $u_i$ being constants.
We have shifted $v$ by a constant to remove the $v^{n-1}$ term. 
This is again equivalent to the standard solution \nref\standard{
P. C. Argyres, M. R. Plesser, and A. D. Shapere, ``The Coulomb Phase Of $N=2$
Supersymmetric QCD,'' hep-th/9505100, Phys. Rev. Lett. {\bf 75} (1995) 1699.}
\nref\other{
A. Hanany and Y. Oz, ``On The Quantum Moduli Space Of Vacua Of $N=2$ Supersymmetric
$SU(N_C)$ Gauge Theories,'' hep-th/9505075, Nucl. Phys. {\bf B 452} (1995) 283.}
\refs{\standard,\other}
 of the $SU(k)$ theory with $N_f$ flavors.  As long as $N_f\not= 2k$,
one can rescale $\tilde t$ and $v$ to set $e=f=1$.  Of course,
shifting $v$ by a constant to eliminate the $v^{k-1}$ term in $B$ will
shift the $m_j$ by a constant.  This is again a familiar part of the solution
of the models.

Of special interest is the case $N_f=2k$ where the beta function vanishes.
In this case, by rescaling $\tilde t$ and $v$, it is possible to remove only
one combination of $e$ and $f$.  The remaining combination is a modulus
of the theory, a coupling constant.  This is as expected: four-dimensional 
quantum
Yang-Mills theory has a dimensionless coupling constant when and only when
the beta function vanishes.  

The coupling constant for $N_f=2k$ is coded into the behavior  of the 
fivebrane for
$z, t\to\infty$.  This behavior, indeed, is a ``constant of the motion'' 
for
finite energy  disturbances of the fivebrane configuration and hence can be
interpreted as a coupling constant in the four-dimensional quantum field 
theory.
The behavior at infinity for $N_f=2k$ is
\eqn\judo{ t\cong \lambda_\pm v^{k},}
where
$\lambda_\pm$ are the two roots of the quadratic equation
\eqn\juppo{ y^2+e y +f=0.}
This follows from the fact that the asymptotic behavior of the equation is
\eqn\asfo{t^2+e(v^k+\dots) t +f(v^{2k}+\dots)=0.}
$y$ can be identified as $t/v^k$.
The fact that the two fivebranes are parallel at infinity -- on both branches
$ t\cong v^{k}$ for $v\to\infty$
-- means that the distance between them has a limit at infinity,
which determines the gauge coupling constant.  

A rescaling of $ t$ or $v$ rescales $\lambda_\pm$ by a common factor,
leaving fixed the function $w=-4\lambda_+\lambda_-/(\lambda_+-\lambda_-)^2$ 
which is
also invariant under exchange of the $\lambda$'s.
This function can be constructed as a product of cross ratios of the
four distinguished points $0,\infty,\lambda_+$ and $\lambda_-$ on the $
y$
plane.  Let ${\cal M}_{0,4;2}$ be the moduli space of the following objects:
a smooth Riemann surface of genus zero with four distinct marked points, two 
of which
(0 and $\infty$) are distinguished and ordered, 
while the others ($\lambda_+$ and $\lambda_-$)
are unordered.  The choice of a value of $w$ (not equal to zero or infinity) 
is the choice of a point in ${\cal M}_{0,4;2}$.  The point $w=1$ is a $\Z_2$
orbifold point in ${\cal M}_{0,4;2}$.

In the conventional description of this theory, one introduces a coupling 
parameter
$\tau$ appropriate near one component of ``infinity'' in ${\cal M}_{0,4;2}$
--  near $w\to 0$ 
(which corresponds for instance to $\lambda_+\to 0$ at fixed $\lambda_-$), 
where the $SU(k)$ gauge theory is weakly coupled.\foot{
The other possible degeneration is $w\to \infty$ ($\lambda_+\to \lambda_-$).  
This
is in $M$ theory the limit of coincident fivebranes, and a weakly coupled
description in four dimensions is not obvious.}  In \standard, the solution
\yuyu\ is expressed in terms of $\tau$.  Near $w=0$ one  has
$w=e^{2\pi i\tau}$; the inverse function $\tau(w)$ is many-valued.  The fact
that the theory depends only on $w$ and not on $\tau$ is from the standpoint 
of 
weak coupling interpreted as the statement that the
theory is invariant under a discrete group of duality transformations.
This group is $\Gamma= \pi_1({\cal M}_{0,4;2})$.
It can be shown that $\Gamma$ is isomorphic to the index three subgroup
of $SL(2,\Z)$ consisting of integral unimodular matrices
\eqn\kkk{\left(\matrix{a & b \cr c & d \cr}\right)}
with $b$ even; this group is usually called $\Gamma_0(2)$.

\bigskip\noindent
{\it The Case Of A Positive Beta Function}

What happens  when the $SU(k)$ gauge theory has positive
beta function, that is for $N_f>2k$?  The fivebrane configuration \yuyu\ still 
describes
something, but what?  The first main point to note is that for $N_f>2k$,
the two fivebranes are parallel near infinity; both branches of \yuyu\ behave
for large $v$ as $\tilde t\sim v^{N_f/2}$.  I interpret this to mean that the 
four-dimensional
theory induced from the branes is conformally invariant at short distances
and flows in the infrared to the $SU(k)$ theory with $N_f$ flavors.

What conformally invariant theory is this?  A key is that
for $N_f\geq 2k+2$, there are additional terms that can be added to \yuyu\
 without changing the
asymptotic behavior at infinity (and cannot be absorbed in redefining $\tilde 
t$ and $v$). 
 Such terms really should be included because
they represent different vacua of the same quantum system.

There are two rather different cases to consider.
If $N_f=2k'$ is even, the general curve with the given behavior at infinity is
\eqn\dogo{\tilde t^2={1\over 4}e'(v^{k'}+\dots)^2-f\prod_{i=1}^{2k'}(v-m_i).}
This describes the the $SU(k')$ theory with $2k'$ flavors, a theory that
is conformally invariant in the ultraviolet and which by suitably
adjusting the parameters 
can reduce in an appropriate limit (taking $e'$ to zero while rescaling $v$ and
some of the other parameters)
to the solution \yuyu\ for the $SU(k)$ theory with $N_f>2k$ flavors.
The $SU(k')$ theory with $2k'$ flavors has of course a conventional Lagrangian
description, valid when the coupling is weak.

The other case is $N_f=2k'+1$, with $k'\geq k$.  
The most general curve with the same asymptotic behavior as \yuyu\ is then
\eqn\odogo{\tilde t^2={1\over 4}e'(v^{k'}+\dots)^2-f\prod_{i=1}^{2k'+1}(v-m_i)
.}
There is no notion of weak coupling here; the asymptotic behavior of
the fivebranes is $\tilde t=\lambda_\pm v^{n'+1/2}$ with 
$\lambda_-=-\lambda_+$ so that
$w$ has the fixed value $1$. (We recall that this is the $\Z_2$ orbifold
point on ${\cal M}_{0,4;2}$.)  \odogo\ describes a
strongly coupled fixed point with no obvious Lagrangian description
and no dimensionless ``coupling
constant,'' roughly along the lines of the fixed point analyzed in
\ref\argdoug{P. Argyres and M. Douglas, ``New Phenomena In $SU(3)$
Supersymmetric Gauge Theory,''
hep-th/9505062, Nucl. Phys. {\bf B448} (1995) 93.}.
  By specializing some parameters, it can flow in the infrared to
the $SU(k)$ theory with $2k'+1$ flavors for any $k<k'$.  

Also, the $SU(k'+1)$ theory with $2k'+2$ flavors can flow in the infrared to
the fixed point just described.  This is done starting with \yuyu\
by taking one mass to infinity while shifting and adjusting the other
variables in an appropriate fashion.

In the rest of this paper, we concentrate on models of zero or negative beta
function.  Along just the above lines, models of positive beta function
can be derived from conventional fixed points like the one underlying
\dogo\ or unconventional ones like the one underlying \odogo; the conventional
and unconventional fixed points are linked by renormalization group flows.

\subsec{Generalization}

Now we will consider a more general model, with a chain of $n+1$
fivebranes labeled from 0 to $n$,
 the $\alpha-1^{th}$ and $\alpha^{th}$  fivebranes, for 
$\alpha=1,\dots,
n$, being connected by $k_\alpha$ fourbranes.  We assume that there are no
semi-infinite fourbranes at either end.\foot{By analogy with the $SU(k)$ theory
with $N_f$ hypermultiplets treated in the last subsection,
semi-infinite fourbranes would be incorporated by taking
the coefficients of $t^{n+1}$ and $t^0$ in the polynomial $P(t,v)$ introduced
below to be polynomials in $v$ of positive degree.  This gives solutions of
models that are actually special cases of models that will be treated in 
section 3.}

The gauge group is thus $\prod_{\alpha=1}^n  SU(k_\alpha)$, and the coefficient
in the one-loop beta function of $SU(k_\alpha)$ is 
\eqn\obo{b_{0,\alpha}= -2k_\alpha+k_{\alpha+1}+k_{\alpha-1}.}
(We understand here $k_{0}=k_{n+1}=0$.)
We will assume $b_{0,\alpha}\leq 0$ for all $\alpha$.  Otherwise,
as in the example just treated, the model is not really a $\prod_\alpha 
SU(k_\alpha)$
gauge theory at short distances but should be interpreted in terms of a 
different 
ultraviolet fixed point.  Note that $\sum_{\alpha=1}^n b_{0,\alpha}<0$ (in fact
$\sum_{\alpha=1}^n b_{0,\alpha}=-k_1-k_n$),
so it is impossible in a model of this type for all beta functions to vanish.
(Models with vanishing beta function can be obtained by including semi-infinite
fourbranes at the ends of the chain, as above, or by other generalizations
made in sections three and four.)

If the position $t_\alpha(v)$ of the 
$\alpha^{th}$ fivebrane, for $\alpha=0,\dots, n$, behaves for large $v$ as
\eqn\iko{t_\alpha(v) \sim h_\alpha v^{a_\alpha}}
with $a_0\geq a_1\geq a_2\dots \geq a_n$ and constants $h_\alpha$,
then from our analysis of the relation of the beta function to 
``bending'' of fivebranes, we have
\eqn\ubu{a_\alpha-a_{\alpha-1}=-b_{0,\alpha},\,\,\,{\rm for}\,\alpha=1,\dots, 
n.}

The fivebrane worldvolume will be described by a curve $F(v,t)=0$,  for some
polynomial $F$.   $F$ will be of degree $n+1$ in $t$ so that for each
$v$ there are $n+1$ roots $t_\alpha(v)$ (already introduced in \iko), 
representing the $v$-dependent
positions of the fivebranes.  $F$ thus has the general form
\eqn\plop{F(t,v)=t^{n+1}+f_1(v)t^n+f_2(v)t^{n-1}+\dots + f_{n}(v)t+1.}
As in the special case considered in section 2.4,
the coefficients of $t^{n+1}$ and $t^0$
are non-zero constants  to ensure the absence
of semi-infinite fourbranes; those coefficients have been set to 1 by scaling
$F$ and $t$.  Alternatively, we can factor $F$ in terms of its roots:
\eqn\ikko{F=\prod_{\alpha=0}^n(t-t_\alpha(v)).}

The fact that the $t^0$ term in \plop\ is independent of $v$ implies that
\eqn\kollo{\sum_{\alpha=0}^na_\alpha=0,}
and this, together with the $n$ equations \ubu, determines the $a_\alpha$ for
$\alpha=0,\dots,n$.  
The solution is in fact
\eqn\hoggog{a_\alpha=k_{\alpha+1}-k_\alpha}
with again $k_{0}=k_{n+1}=0$.

If the degree of a polynomial $f(v)$ is denoted by $[f]$, then
the factorization \ikko\ and asymptotic behavior \iko\ imply that
\eqn\loppo{[f_1]=a_0, \,\,[f_2]=a_0+a_1,\,\,[f_3]=a_0+a_1+a_2,\dots.}
Together with \hoggog, this implies simply
\eqn\kimbo{[f_\alpha]=k_{\alpha}, \,{\rm for}\,\alpha=1,\dots, n.}
If we rename $f_\alpha$ as $p_{k_{\alpha}}(v)$, where the
subscript now equals  the degree of a polynomial, then the polynomial 
$F(t,v)$ takes the form
\eqn\hombo{F(t,v)=t^{n+1}+p_{k_1}(v)t^n+p_{k_2}(v)t^{n-1}+\dots 
+p_{k_n}(v)t+1.}

The curve $F(t,v)=0$ thus describes the solution of the model with gauge
group $\prod_{\alpha=1}^nSU(k_\alpha)$ and hypermultiplets in the 
representation
$\sum_{\alpha=1}^{n-1}(\k_\alpha,\bar\k_{\alpha+1}).$  
The fact that the coefficient of $t^i$, for $1\leq i\leq n$, is a polynomial
of degree $k_i$ in $v$ has a clear intuitive interpretation: the zeroes
of $p_{k_\alpha}(v)$ are the positions of the $k_\alpha$ fourbranes that 
stretch between
the $\alpha^{th}$ and $\alpha+1^{th}$ fivebrane.  

The polynomial $p_{k_\alpha}$ has the form
\eqn\mimbo{p_{k_\alpha}(v)=c_{\alpha,0}v^{k_\alpha}+c_{\alpha,1}v^{k_\alpha-1}+
  c_{\alpha,2}v^{k_\alpha-2}+  \dots +c_{\alpha,k_\alpha}.}  
The leading coefficients $c_{\alpha,0}$ determine the asymptotic positions of
the fivebranes for $v\to \infty$, or more precisely the constants $h_\alpha$ 
in \iko.
In fact by comparing
the factorization $F(t,v)=\prod_\alpha(t-t_\alpha(v))=\prod_\alpha(t-h_\alpha
t^{v_\alpha}+O(t^{v_\alpha-1}))$ to the series \hombo\ one can express the 
$h_\alpha$ in terms of
 the $c_{0,\alpha}$. 

The $h_\alpha$ determine the constant terms in 
the asymptotic freedom formula
\eqn\pllo{-i\tau_\alpha=-b_{0,\alpha}\ln v+{\rm constant}}
for the large $v$ behavior of the inverses of the
effective gauge couplings.   
Thus, the $c_{\alpha,0}$'s should be identified with the gauge coupling 
constants.
Of course, one combination of the $c_{\alpha,0}$'s can
be eliminated by rescaling the $v$'s; this can be interpreted as a 
renormalization
group transformation via which (as the beta function coefficients
$b_{0,\alpha}$ are not all zero) one
coupling constant can be eliminated.

In particular, the $c_{\alpha,0}$ are constants that parametrize the choice
of a quantum system, not order parameters that determine the choice of a vacuum
in a fixed quantum system.
The $c_{\alpha,1}$ are likewise constants, according to \boggo; they
determine the hypermultiplet bare masses.  (One of the $c_{\alpha,1}$ can
be removed by adding a constant to $v$; in fact there are $n$ $c_{\alpha,1}$'s 
and
only $n-1$ hypermultiplet bare masses.)  The $c_{\alpha,s}$ for $s=2,\dots,
k_\alpha$ are the order parameters 
on the Coulomb branch of the $SU(k_\alpha)$ factor of the gauge group.

\newsec{Models With Sixbranes}

\subsec{Preliminaries}

\midinsert
\centerline{\psfig{figure=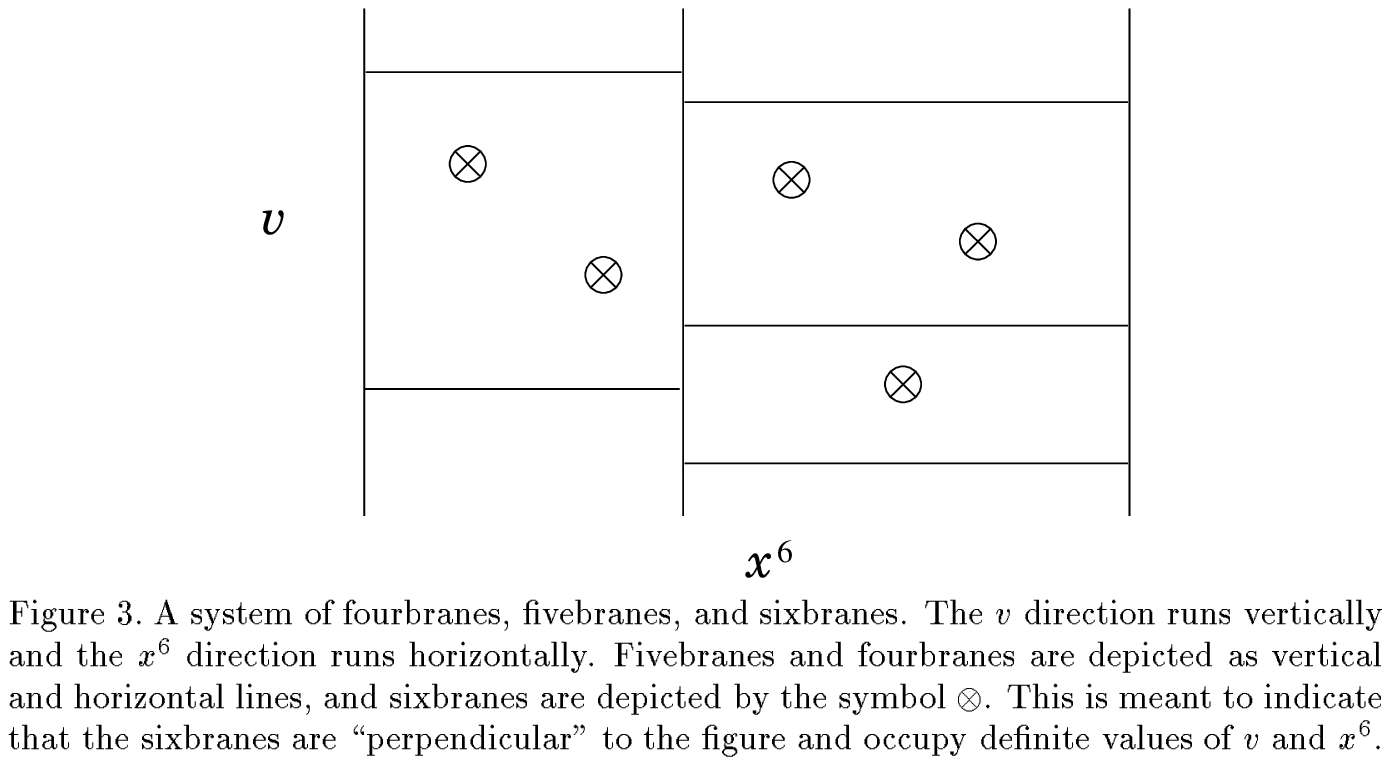,width=4.5in}}
\bigskip\bigskip
\endinsert
The goal in the present section is to incorporate sixbranes in the models
of the previous section. 
The sixbranes will enter just like the
D fivebranes in \hw\ and for some purposes can be analyzed quite similarly.

Thus we consider again the familiar
 chain of $n+1$ fivebranes, labeled from 0 to $n$, with  $k_\alpha$ fourbranes
stretched between the $\alpha-1^{th}$ and $\alpha^{th}$ fivebranes, for 
$\alpha=1,\dots,n$.
But now we place $d_\alpha$ sixbranes between the $\alpha-1^{th}$ and 
$\alpha^{th}$ fivebranes,
for $\alpha=1,\dots,n$.  A special case is sketched in figure 3.
In the coordinates introduced at the beginning of 
section
two, each sixbrane is  located at definite values
of $x^4,x^5$, and $x^6$ and has a world-volume that is parametrized
by arbitrary values of $x^0,x^1,\dots, x^3$ and $x^7,x^8,$ and $x^9$.

Given what was said in section two and in \hw,  the interpretation of the 
resulting
model as a four-dimensional gauge theory is clear.  The gauge group
is $\prod_{\alpha=1}^n SU(k_\alpha)$.  
The hypermultiplets consist of the $(\k_\alpha,\bar \k_{\alpha+1})$ 
hypermultiplets
that were present without the sixbranes, plus additional hypermultiplets
that become massless whenever a fourbrane meets a sixbrane.  As in \hw, these
additional hypermultiplets transform in $d_\alpha$ copies of the fundamental 
representation
of $SU(k_\alpha)$, for each $\alpha$.  The bare masses of these 
hypermultiplets are
determined by the positions of the sixbranes in $v=x^4+ix^5$.  As in     \hw,
the positions of the sixbranes in $x^6$ decouple from many aspects of
the low energy four-dimensional physics.

One difference from section two is that (even without semi-infinite 
fourbranes) there
are many models with vanishing beta function.  In fact, for each choice of     
 $k_\alpha$
such that the models considered in section two had all beta functions zero or
negative, there is upon inclusion of sixbranes a unique
choice of the $d_\alpha$ for which the beta functions all vanish,
namely
\eqn\jolly{d_\alpha=2k_\alpha-k_{\alpha+1}-k_{\alpha-1}}
(where we understand that $k_{0}=k_{n+1}=0$).
 By solving all these
models, we will get a much larger class of solved $\N=2$ models with zero beta
function than has existed hitherto.  For each such model,
one expects to find a non-perturbative duality group generalizing the duality 
group
$SL(2,\Z)$ of four-dimensional $\N=4$ super Yang-Mills theory.  
{}From the solutions we will get, the duality groups turn out to be
as follows. Let ${\cal M}_{0,n+3;2}$ be the moduli space of objects of the 
following
kind: a smooth Riemann surface of genus zero with $n+3$ marked points, two
of which are distinguished and ordered while the other $n+1$ are unordered.  
Then the duality group of a model with $n+1$ fivebranes is the fundamental
group $\pi_1({\cal M}_{0,n+3;2})$.  One can think roughly of the genus zero 
Riemann
surface in the definition of ${\cal M}_{0,n+3;2}$
as being parametrized by the variable $t$ of section two, with the
marked points being 0, $\infty$, and the positions of the $n+1$ fivebranes.

In contrast to section two, we would gain nothing essentially new by 
incorporating semi-infinite fourbranes at the two ends of the chain.  This
gives hypermultiplets in the fundamental representation of the groups 
$SU(k_1)$ and
$SU(k_n)$ that are supported at the ends of the chain; we will anyway generate
an arbitrary number of such hypermultiplets via sixbranes.  Another 
generalization
that would give nothing essentially new would be to include fourbranes that 
connect
fivebranes to sixbranes.  Using a mechanism considered in \hw, one can by 
moving the sixbranes in the
$x^6$ direction reduce to the  case that all fourbranes end on fivebranes.
One could also add sixbranes to the left or to the right of all fivebranes.
In fact, we will see how this generalization can be incorporated in the 
formulas.
In the absence of fourbranes ending on them,  sixbranes that are to the left
or right of everything else simply decouple
from the low energy four-dimensional physics.  

Another generalization is to consider fourbranes that end
on sixbranes at both ends.  As in \hw, such a fourbrane supports a 
four-dimensional
hypermultiplet, not a vector multiplet, and configurations containing such
fourbranes must be included to describe Higgs branches (and mixed 
Coulomb-Higgs branches)
of these theories.  We will briefly discuss the Higgs branches in section 3.5.

\subsec{Interpretation In $M$ Theory}

Since our basic technique is to interpret Type IIA brane configurations in $M$
theory, we need to know how to interpret the Type IIA sixbrane in $M$ theory.
This was first done in \ref\townold{P. Townsend, ``The
Eleven-Dimensional Supermembrane Revisited,'' Phys. Lett. {\bf B350} (1995) 
184.}.

Consider $M$ theory on $\R^{10}\times \S^1$.  This is equivalent to Type IIA
on $\R^{10}$, with the $U(1)$ gauge symmetry of Type IIA being associated in 
$M$
theory with the rotations of the $\S^1$.  States that have momentum in the
$\S^1$ direction are electrically charged with respect to this $U(1)$ gauge 
field
and are interpreted in Type IIA as Dirichlet zerobranes.
The sixbrane is the electric-magnetic dual of the zerobrane, so it is 
magnetically
charged with respect to this same $U(1)$. 

The basic object that is magnetically charged with respect to this $U(1)$ 
is the ``Kaluza-Klein monopole'' or Taub-NUT space.
This is derived from a hyper-Kahler solution of the four-dimensional 
Einstein equations.   The metric is asymptotically flat, and the space-time
looks near infinity like a non-trivial $\S^1$ bundle over $\R^3$.  The 
Kaluza-Klein
magnetic charge is given by the 
twisting of the $\S^1$ bundle, which is incorporated in the formula given below
by the appearance of the Dirac monopole potential.

\def\r{{\bf r}}
Using conventions of \ref\gibbons{G. W. Gibbons and P. Rychenkova,
``HyperK\"ahler Quotient Construction Of BPS Monopole Moduli Spaces,''
hep-th/9608085.} adapted to the notation
of the present paper, if we define a three-vector
$\vec\r=(x^4,x^5,x^6)R$, and set $r=|\vec\r|$ and  $\tau=x^{10}/R$, 
then the Taub-NUT metric is 
\eqn\uffo{ds^2={1\over 4}\left({1\over r}+{1\over R^2}\right)d\vec\r^2
+{1\over 4}\left({1\over r}+{1\over R^2}\right)^{-1}(d\tau+\vec\omega\cdot 
d\vec
\r)^2.}
Here $\vec\omega$ is the Dirac monopole potential (which one can identify
locally as a one-form obeying $\vec\nabla\times \vec \omega =\vec \nabla(1/r)$).

To construct a sixbrane on $\R^{10}\times \S^1$, we simply take the product
of the metric \uffo\ with a flat metric on $\R^{7}$ (the coordinates on
$\R^7$ being $x^0,\dots,x^3$ and $x^7,\dots,x^9$).  We will be interested
in the case of many parallel sixbranes, which is described by the 
multi-Taub-NUT metric \ref\hawking{S. Hawking, ``Gravitational Instantons,''
Phys. Lett. {\bf 60A} (1977) 81.}:
\eqn\turgo{ds^2={V\over 4}d\vec\r^2+{V^{-1}\over 4}(d\tau+\vec
\omega\cdot d\vec r)^2,}
where now
\eqn\toompo{V=1+\sum_{a=1}^d{1\over |\r-{\bf x}_a|}}
and $\vec\nabla\times \omega=\vec\nabla V$.  This describes a configuration
of $d$ parallel sixbranes, whose positions are the ${\bf x}_a$.

The reason that by going to eleven dimensions we will get some simplification
in the study of sixbranes is that, in contrast to the ten-dimensional low 
energy field
theory
in which the sixbrane core is singular, in $M$ theory the sixbrane
configuration is described by the multi-Taub-NUT metric \toompo, which is
complete and smooth (as long as the $ {\bf x}_a$ are distinct).  This
elimination of the sixbrane singularity was in fact emphasized in \townold.
In going from $M$ theory to Type IIA, one reduces from eleven to ten dimensions
by dividing by the action of the vector field $\partial/\partial\tau$.
This produces singularities at $d$ points at which $\partial/\partial\tau$
vanishes; those $d$ points are interpreted in Type IIA as 
positions of sixbranes.  In general in physics, appearance of singularities
in a long wavelength description means that to understand the behavior of a 
system
one needs more information.  The fact that the sixbrane singularity is
eliminated in going to $M$ theory means that, if the radius $R$ of the
$x^{10}$ circle is big,\foot{We recall that we can assume this radius to be big
since it corresponds to an ``irrelevant'' parameter in the field
theory.}  the $M$ theory can be treated via low energy 
supergravity.  This is just analogous to what happened in section 2; the
singularity of Type IIA fourbranes ending on fivebranes was eliminated
upon going to $M$ theory, as a result of which low energy supergravity
was an adequate approximation.  
The net effect is that unlike either long wavelength ten-dimensional field 
theory
or conformal field theory, the long wavelength eleven-dimensional field theory
is an adequate approximation for the problem.  

In this paper we will really not use the hyper-Kahler metric of the
multi-Taub-NUT space, but only the structure (or more exactly one of the structures)
as a complex manifold.   If as before we set $v=x^4+ix^5$, then
in one of its complex structures the multi-Taub-NUT space can be described by 
the equation
\eqn\juddo{yz=\prod_{a=1}^d(v-e_a)}
in a space $\C^3$ with three complex coordinates $y,z$, and $v$.
Here $e_a$ are the positions of the sixbranes projected to the complex $v$ 
plane.
Note that \juddo\ admits the
$\C^*$ action
\eqn\uddo{y\to \lambda y,\,\,\,z\to \lambda^{-1} z,}
which is the complexification of the $U(1)$ symmetry of \turgo\ that is 
generated
by $\partial/\partial \tau$.
For the special case that there are no fivebranes, this $\C^*$ corresponds
to the transformation $t\to \lambda t$ where $t=\exp\left(-(x^6+ix^{10})/R\right)$.
Hence very roughly, 
for large $y$ with fixed or small $z$, 
$y$ corresponds to $t$ and for large $z$ with fixed
or small $y$, $z$ corresponds to $t^{-1}$.  
(As there is a symmetry exchanging $y$ and $z$, their roles could be reversed
in these assertions.)

In  section 3.6, we will use the approach of \gibbons\ to show
that the multi-Taub-NUT space is equivalent as a complex manifold to \juddo.  
The formulas in section 3.6 can also be used to make the asymptotic
identification of
$y$ and $z$ with $t$ and $t^{-1}$ more precise.
For now, we note the following facts, which may orient the reader.
When all $e_a$ are coincident at, say, $v=0$, \juddo\ reduces
to the $A_{n-1}$ singularity $yz=v^n$.  A system of  parallel and coincident
sixbranes in Type IIA generates a $U(n)$ gauge symmetry; the $A_{n-1}$ 
singularity is the mechanism by which such enhanced gauge symmetry 
appears in the $M$ theory description.  In general, \juddo\ describes
the unfolding of the $A_{n-1}$ singularity.

The complex structure \juddo\ does not uniquely fix the hyper-Kahler metric,
not even the behavior of the metric at infinity.  The {\it same}
complex manifold  \juddo\ admits a family  of ``asymptotically locally
Euclidean'' (ALE) metrics, which look at infinity like $\C^2/{\bf Z}_n$.
(They are given by the same formula \turgo, but with a somewhat different 
choice
of $V$.)  The metrics \turgo\ are not ALE but are  ``asymptotically locally 
flat'' (ALF).

Even if one asks for ALF behavior at infinity, the hyper-Kahler metric
involves parameters that do not appear in \juddo.  The hyper-Kahler metric 
\turgo\
depends on the positions $\vec {\bf x}_a$ of the sixbranes, while in \juddo\
one sees only the projections $e_a$ of those positions to the $v$ plane.
{}From the point of view of the complex structure that is exhibited in \juddo,
the  $x^6$ component of the sixbrane positions is coded in the Kahler class
of the metric \turgo.

In studying the Coulomb branch of $\N=2$ models, we will really need only the
complex structure \juddo; the $x^6$ positions of sixbranes will be irrelevant.
This is analogous to the fact  that in studying the Coulomb branch
of $N=4$ models in three dimensions by methods of \hw, the $x^6$ positions
of Dirichlet fivebranes are irrelevant.  As that example suggests,
the $x^6$ positions are relevant for understanding the Higgs branches
of these models.  

In one respect, the description \juddo\ of the complex structure is
misleading.  Whenever $e_a=e_b$ for some $a$ and $b$, the complex manifold
\juddo\ gets a singularity.  The hyper-Kahler metric, however, becomes
singular only if two sixbranes have equal positions in $x^6$ and not only
in $v$.  When two sixbranes have the same position in $v$ but not in $x^6$,
the singular complex manifold \juddo\ must be replaced by a smooth one that
is obtained by blowing up the singularities, replacing each $A_k$ singularity
by a configuration of $k$ curves of genus zero.  This subtlety will be 
important
when, and only when, we briefly examine the Higgs branches of these models.

\subsec{$\N=2$ Supersymmetric QCD Revisited}

\midinsert
\centerline{\psfig{figure=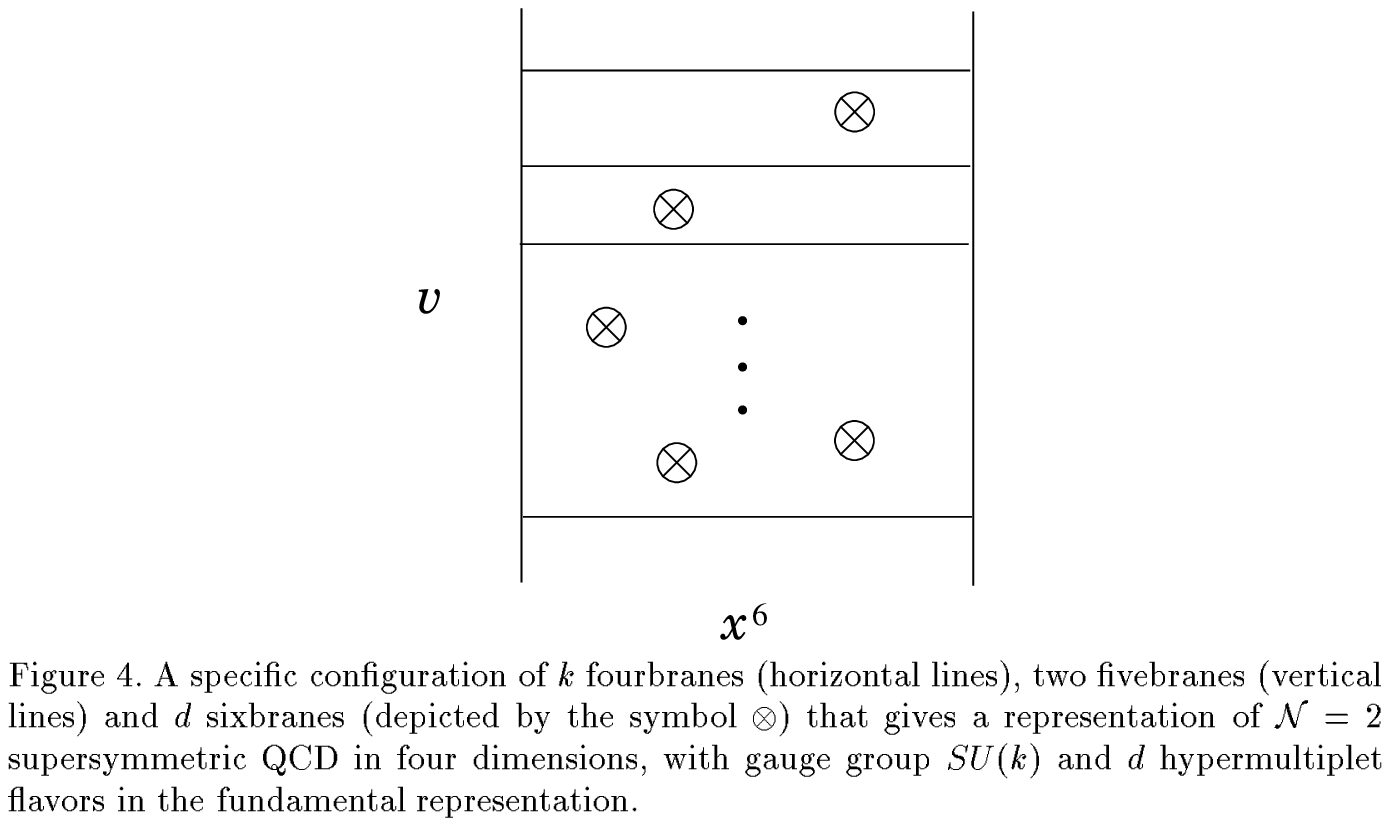,width=4.5in}}
\bigskip\bigskip
\endinsert
Now we want to solve for the Coulomb branch of a model that is
constructed in terms of Type IIA via a configuration of fourbranes,
fivebranes, and sixbranes.  The only change from section 2 is that
to incorporate sixbranes
we must replace  $Q=\R^3\times \S^1$, in which
the $M$ theory   fivebrane propagated
in section 2, by the multi-Taub-NUT space $\tilde Q$ that
was just introduced.  We write the defining equation of $\tilde Q$ as
\eqn\uxx{yz=P(v),}
with $P(v)=\prod_{a=1}^d(v-e_a)$.  
Type IIA fourbranes and fivebranes are described as before by a complex curve
$\Sigma$ in $\tilde Q$. 
$\Sigma$ will be described by an equation $F(y,v)=0$.  Note that we can assume
that $F$ is independent of $z$, because $z$ could be eliminated via $z=P(v)/y$.
 
For our first attempt to understand the combined system of fourbranes, 
fivebranes, 
and sixbranes, we consider the example in figure 4 of two parallel 
fivebranes
connected by $k$ fourbranes, with
$d$ sixbranes between them.  We assume that there are no semi-infinite
fourbranes extending to the left or right of the figure.  This configuration
should correspond to $\N=2$ supersymmetric QCD, that is
to an $SU(k)$ gauge theory with $d$ hypermultiplets
in the fundamental representation of $SU(k)$.  

As in section 2, the fact that there are two fivebranes means that
the equation $F(y,v)=0$, regarded as an equation in $y$ for fixed $v$,
has generically two roots.  Thus, $F$ is quadratic in $y$ and has the general
form
\eqn\mumbo{A(v)y^2+B(v)y+C(v)=0.}
By clearing denominators and dividing by common factors, we can assume
that $A,B$, and $C$ are relatively prime polynomials.

Now we must interpret the statement that there are no semi-infinite fourbranes.
This means, as in section 2, that it is impossible for $y$ or $z$ (which
correspond roughly to $t$ and $t^{-1}$ in the notation of section 2) 
to go to infinity at a finite value of $v$.  The requirement that $y$ never
diverges at finite $v$ means that -- if $A,B$, and $C$ are understood to
have no common factors -- $A(v)$ is a constant, which we can take to equal 1.
So the defining equation of $\Sigma$ reduces to
\eqn\pumbo{y^2+B(v)y+C(v)=0.}
Now let us express this in terms of $z=P(v)/y$.  We get
\eqn\umbo{C(v)z^2+B(v)P(v) z+P(v)^2=0.}
$z$ will diverge at zeroes of $C$ unless both $BP$ and $P^2$ are
divisible by $C$.  Such divergence would represent the existence of a 
semi-infinite
fourbrane.

In particular, the absence of semi-infinite fourbranes implies that $P^2$
is divisible by $C$.  So any zero of $C$ is a zero of $P$, that is,
it is one of the $e_a$.  Moreover, in the generic case that the $e_a$ are
distinct, each $e_a$ can appear as a root of
$C$ with multiplicity at most two.  Thus, we can label the $e_a$
in such a way that $e_a$ is a root of $C$ with multiplicity 2 for
$a\leq i_0$, of multiplicity $1$ for $i_0<a\leq i_1$, and of multiplicity
0 for $a>i_1$. 
We then have
\eqn\hurry{C=f\prod_{a=1}^{i_0}(v-e_a)^2\prod_{b=i_0+1}^{i_1} (v-e_b)}
with some non-zero complex constant $f$.
The requirement that $BP$ should be divisible 
by $C$ now implies that the $e_a$ of $a\leq i_0$ are roots of $B$, so
\eqn\hulf{B(v)  =\tilde B(v) \prod_{a\leq i_0}(v-e_a)}
for some polynomial $\tilde B$. 

The equation \mumbo\ now reduces to
\eqn\iffogo{y^2+\tilde B(v)\prod_{a\leq i_0}(v-e_a)y +f\prod_{a\leq 
i_0}(v-e_a)^2
\prod_{b=i_0+1}^{i_1}(v-e_b)
 =0.}
In terms of $\tilde y=y/\prod_{a\leq i_0}(v-e_a)$, this is
\eqn\ikkogo{\tilde y^2+\tilde B(v) \tilde y + f\prod_{a=i_0+1}^{i_1} (v-e_a) 
=0.}
If $\tilde B(v)$ is a polynomial of degree $k$, this is (for $i_1-i_0\leq 2k$;
otherwise as at the end of section 2 one encounters a new ultraviolet
fixed point) the familiar solution of
the $SU(k)$ gauge theory with $i_1-i_0$ flavors in the fundamental 
representation,
written in the same form in which it appeared in section 2.
The $e_a$ with $a\leq i_0$ or $a>i_1$ have decoupled from the gauge
theory.

This suggests the following interpretation: the sixbranes  with $a\leq i_0$ are
to the left of all fivebranes, the sixbranes with $i_0+1\leq a\leq i_1$ are
between the two fivebranes, and the sixbranes with $a>i_1$ are to the right
of all fivebranes.  If so then (in the absence of fourbranes ending on the 
sixbranes)
the sixbranes with $a\leq i_0$ or $a>i_1$ would be decoupled from the 
four-dimensional gauge theory, and the number of hypermultiplet copies of the
fundamental representation of $SU(k)$ would be $i_1-i_0$, as we have just seen.
We will now justify that interpretation.

\bigskip\noindent{\it Interpretation  Of $i_0$ and $i_1$}

The manifold $\tilde Q$ defined by $yz=P(v)$ maps to the complex $v$ plane,
by forgetting $y$ and $z$.  Let $Q_v$ be the fiber of this map for a given 
value of
$v$.  For generic $v$, the fiber is a copy of $\C^*$.  Indeed, whenever 
$P(v)\not= 0$,
the fiber $Q_v$, defined by
\eqn\imm{yz=P(v),}
is a copy of $\C^*$ (the complex $y$ plane with $y=0$ deleted).  
This copy of $\C^*$ is actually an orbit of the $\C^*$ action \uddo\ on 
$\tilde Q$.

We recall from section 3.2 that if $z$ or $y$ is large with the other fixed, then
the asymptotic relation between $z$, $y$, and $t=\exp\left(-(x^6+ix^{10})/R\right)$
is $y\cong t$ or $z\cong t^{-1}$.  $t\to 0$ means large $x^6$,
which we call ``being on the right''; $t\to \infty$ means $x^6\to -\infty$,
which we call ``being on the left.'' Thus $z$ much larger than $y$ or vice-versa
corresponds to being on the right or on the left in $x^6$.

The surface $\Sigma$ is defined by an equation $F(y,v)=0$ where $F$ is 
quadratic
in $y$; it intersects each $Q_v$ in two points.  ($Q_v$ is not complete, but
we have chosen $F$ so that no root goes to  $y=\infty$ or $z=\infty$ for $v$ such
that $P(v)\not= 0$.)   These are
the two points with five-branes, for the given value of $v$.

Now consider the special fibers with $F(v)=0$.  This means that for some $a$, 
$v$ is
equal to $e_a$, 
the position in the $v$ plane of the $a^{th}$ sixbrane.  The fiber $F_v$ is 
for such $v$
defined by
\eqn\kimob{yz=0,}
and is a union of two components $C_v$ and $C_v'$ with, respectively, $z=0$ and
$y=0$.  The total number of intersection points of $\Sigma$ with $F_v$ is still
2, but some intersections lie on $C_v$ and some lie on $C_v'$.  
Without
passing through any singularity, we can go to the case that the intersections
on $C_v$ are at large $y$ and those on $C_v'$ are at large $z$.
Hence, fivebranes that correspond to intersections with
$C_v$ are to the left of the $a^{th}$ sixbrane ($y$ is much bigger than $z$ so
they are at a smaller value
of $x^6$) and fivebranes that correspond to intersections with $C_v'$ are to 
the right
of the $a^{th}$ sixbrane (they are at a larger value of $x^6$).

The intersection points on $C_v$ are the zeroes of \iffogo\ which as $v\to e_a$
do not go to $y=0$.  
The intersection points on $C_v'$ are likewise the zeroes of that polynomial
that {\it do} vanish as $v\to e_a$.
The number of such intersections with $C_v'$ is  
two if $a\leq i_0$, one if $i_0+1\leq a\leq i_1$, and zero otherwise.
This confirms that the number of sixbranes to the left of both fivebranes is
$i_0$, the number which are to the left of one and to the right of the other is
$i_1-i_0$, and the number which are to the right of both is $i_1$.

\subsec{Generalization}

We will now use similar methods to solve for the Coulomb branch of a more 
general
model with $n+1$ fivebranes, joined in a  similar way by fourbranes and
with sixbranes between them. 

The curve $\Sigma$ will now be
defined by the vanishing of a polynomial $F(y,v)$ that is of degree
$n+1$ in $y$:
\eqn\koko{y^{n+1} +A_1(v)y^{n}+A_2(v)y^{n-1}+\dots +A_{n+1}(v)=0.}
The $A_\alpha(v)$ are polynomials in $v$.  We assume that there are no 
semi-infinite
fourbranes and therefore have set the coefficient of $y^{n+1}$ to 1.
Substituting $y=P(v)/z$, we get
\eqn\boko{A_{n+1}z^{n+1}+A_nPz^n+A_{n-1}P^2z^{n-1}+\dots + P^{n+1}=0.}
Hence absence of semi-infinite fourbranes implies that 
$A_\alpha P^{n+1-\alpha}$ is divisible by $A_{n+1}$ for all $\alpha$ 
with $0\leq \alpha \leq n$.  (In this assertion we understand $A_0=1.$)
In particular, $P^{n+1}$ is divisible by $A_{n+1}$.

It follows that all zeroes of $A_{n+1}$ are zeroes of $P$, and occur (if the 
$e_a$ are distinct) 
with multiplicity at most $n+1$.  As in the example  considered
before, zeroes of $P$ that occur as zeroes of $A_{n+1}$ with multiplicity
0 or $n+1$ make no essential contribution (they correspond to sixbranes
that are to the left or the right of everything else and can be omitted).
So we will assume that all zeroes of $P$ occur as zeroes of $A_{n+1}$ with
some multiplicity between $1$ and $n$.  
There are therefore integers $i_0,i_1,\dots, i_{n}$ with $i_0=0\leq i_1\leq 
i_2\leq
\dots \leq i_{n-1}\leq i_{n}=n$ such that
if for $1\leq s\leq n$
\eqn\uggu{J_s=\prod_{a=i_{s-1}+1}^{i_s}(v-e_a)}
then
\eqn\mimbo{A_{n+1}=f\prod_{s=1}^nJ_s^{n+1-s}}
with $f$ a constant.
By an argument along the lines given at the end of section 3.3, 
we can interpret $i_\alpha$ as the number of sixbranes to the left of the
$\alpha^{th}$ fivebrane.
So $d_\alpha=i_{\alpha}-i_{\alpha-1}$ is the number of sixbranes between
the $\alpha-1^{th}$ and $\alpha^{th}$ fivebranes.  
The number of hypermultiplets in the fundamental representation of
the $\alpha^{th}$ factor
of the gauge group will hence be $d_\alpha$.

The requirement that $A_\alpha P^{n+1-\alpha}$ is divisible by $A_{n+1}$ is 
then
equivalent to the statement
\eqn\jujitsu{A_\alpha=g_\alpha(v) \prod_{s=1}^{\alpha-1} J_s^{\alpha-s}}
with some polynomial $g_\alpha(v)$.
We interpret $g_\alpha(v)$ as containing the order parameters for the
$\alpha^{th}$ factor of the gauge group.  So if $g_\alpha(v)$ is of degree 
$k_\alpha$, then the gauge group is 
\eqn\guffy{G=\prod_{\alpha=1}^n SU(k_\alpha).}
The hypermultiplet spectrum consists of the usual $(\k_\alpha,\bar\k_{\alpha+1}
)$
representations plus $d_\alpha$ copies of the fundamental representation
of $SU(k_\alpha)$.

The curve describing the solution of this theory should thus be
\eqn\ohuffy{\eqalign{y^{n+1}+&g_1(v)y^n+g_2(v) J_1(v) y^{n-1} 
+g_3(v) J_1(v)^2J_2(v)y^{n-2} \cr &+\dots +g_\alpha(v) \prod_{s=1}^{\alpha-1}
J_s^{\alpha-s} \cdot y^{n+1-\alpha} +\dots +f
\prod_{s=1}^nJ_s^{n+1-s}=0.\cr }} 

This of course reduces in the absence of sixbranes to the solution found in 
\hombo;
it likewise gives back the standard solution of $\N=2$ supersymmetric QCD when 
there
are precisely two fivebranes.  
As a further check, let us examine the condition on the $d_\alpha$ and the 
$k_\alpha$
under which the beta function vanishes.  Note that the coefficient of $y^n$
is of degree $v^{k_1}$.  All fivebranes
will be parallel at large $v$, and the beta function will vanish,
if the coefficient of $y^{n+1-m}$ is of
order $v^{mk_1}$ for $m=1,\dots,n+1$.  Those conditions can be evaluated to 
give
$k_2+d_1=2k_1$, $k_3+k_1+d_2=2k_2$, and so on -- the standard conditions for 
vanishing
beta function of the gauge theory. 

In this case of vanishing beta function, 
let the polynomials $g_\alpha(v)$ be of the form $g_\alpha(v)=h_\alpha 
v^{k_\alpha}
+O(v^{k_\alpha-1})$.  Then the asymptotic behavior of the roots of \ohuffy\
(regarded as an equation for $y$) is $y\sim \lambda_i v^{k_1}$, where
the $\lambda_i$ are the roots of the polynomial equation
\eqn\ororgan{x^{n+1}+h_1x^n+h_2x^{n-1}+\dots +h_nx+f=0.}
On the $x$ plane, there are $n+3$ distinguished points, namely 0, $\infty$, 
and the
$\lambda_i$.  The $\lambda_i$ are of course defined only up to permutation and
(as one could rescale $y$ and $x$)
up to multiplication by a common complex scalar. 
A choice of the $\lambda_i$, modulo those equivalences, determines the 
asymptotic
distances between fivebranes and hence the bare gauge coupling constants.
The same choice also determines a point in the moduli
space ${\cal M}_{0,n+3;2}$ that was introduced in section 3.1.
In any description by a Lagrangian field theory with coupling parameters
$\tau_i$, the fundamental group $\pi_1({\cal M}_{0,n+3;2})$ would be 
interpreted
as the group of discrete duality symmetries.  

\subsec{Higgs Branches}

In this subsection, we will sketch how the transition to a Higgs
branch (or a mixed Higgs-Coulomb branch) can be described from the present point of
view.

\midinsert
\centerline{\psfig{figure=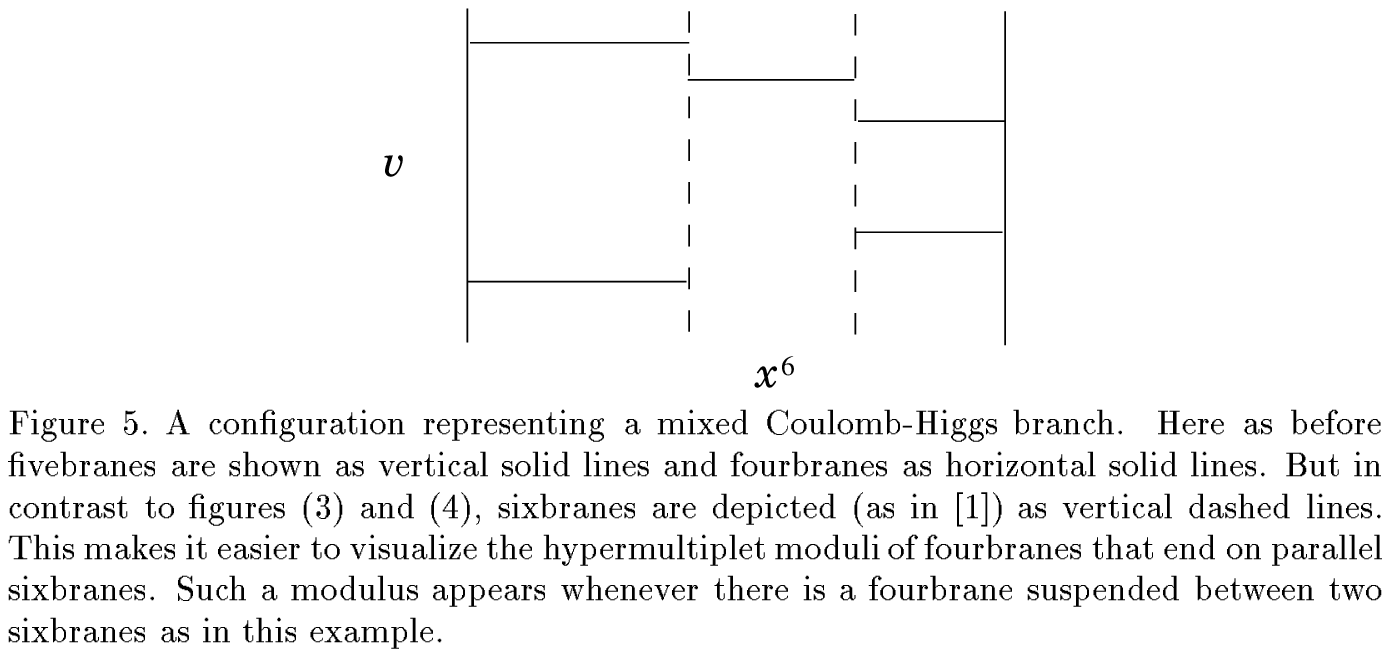,width=4.5in}}
\bigskip\bigskip
\endinsert
We recall that the transition to a Higgs branch is a process in which the
genus of $\Sigma$ drops by one (or more) and a transition
is made to a new branch of vacua in which there are massless hypermultiplets.
In terms of Type IIA brane diagrams, massless hypermultiplets result (as
in \hw) from fourbranes suspended between fivebranes, a configuration
shown  in figure 5.

For a transition to a Higgs branch to occur, it is necessary for two 
hypermultiplet bare masses to become equal.  From the present point of view,
this means that that the positions of two sixbranes in $v$ become equal.
It is {\it not} necessary for the two sixbranes to have equal positions in
$x^6$.  In fact, the semiclassical brane diagram of figure 5 cannot be
drawn if the $x^6$ values of the sixbranes are equal. 

The hypermultiplet bare masses are the roots of $P(v)=\prod_a(v-e_a)$.  
We therefore want to consider the case that two $e_a$ are coincident at,
say, the origin.  The other $e_a$ will play no material role, and we may as
well take the case of only two sixbranes.  So we take $P(v)=v^2$.
The equation $yz=P(v)$ is in this case
\eqn\ollop{xy=v^2,}
and describes a manifold $Q_0$ which
has a singularity at the point $P$ with coordinates $x=y=z=0$.

We recall, however, from the discussion in section 3.2
that in case two sixbranes coincide in $v$ but not in $x^6$, such a singularity
should be blown up.  Thus, the multi-Taub-NUT
manifold $\tilde Q$ does not coincide with $Q_0$, but 
is a smooth surface obtained by blowing up the singularity in $Q_0$.
In the blow-up, $P$ is replaced by a smooth curve
$C$ of genus zero.

Now we consider a curve $\Sigma$ in $\tilde Q$ (or $Q_0$) representing
a point on the Coulomb branch of one of the models considered in this section.
Let $g$ be the generic genus of $\Sigma$.  Nothing essential will be lost if we 
consider the case of supersymmetric QCD -- two fivebranes; gauge group $SU(n)$.  
So $\Sigma$ is defined by a curve of the form
\eqn\uuu{y^2+By +fv^2=0.}
Nothing of interest will happen unless  
$\Sigma$ passes through the singular point $y=z=v=0$.  That is so if and only if
$B$ vanishes at $v=0$ (if $B$ is non-zero at $v=0$ then either $y$ is non-zero
for $v\to 0$, or $y\sim v^2$ for $v\to
0$ and $z$ is non-vanishing at $v=0$), 
so generically $B=bv+O(v^2)$ with a non-zero constant $b$.

So near $P$, $\Sigma$ looks like
\eqn\vuu{y^2+bvy+fv^2=0.}
This curve has a singularity at $y=v=0$.  In fact, the quadratic polynomial
$y^2+bvy+fv^2$ has a factorization as $(y+\gamma v)(y+\gamma' v)$.  Generically,
the two factors correspond, near $P$, to two branches of $\Sigma$ that meet
``tranversely'' at $P$, giving the singularity.
The genus of $\Sigma$ drops by one when this singularity appears.  So $\Sigma$
now has genus $g-1$.  

We actually want to consider the case in which the two sixbranes are not
coincident in $x^6$, so we must consider the curve 
defined by \uuu\ not in the singular manifold 
$Q_0$ but in its smooth resolution $\tilde Q$.  This curve has two components.
One is a {\it smooth} curve $\Sigma'$ of genus $g-1$ and the other
is a copy of the genus zero curve $C$ in $\tilde Q$ that is obtained by
the blowup of $P$.  $\Sigma'$ is smooth (generically) because after the blowup the
two branches $y+\gamma v=0 $ and $y+\gamma' v=0$ of $\Sigma$ no longer meet.
A copy of $C$ is present because the polynomial $y^2+By+v^2$ vanishes
on $P$ and hence (when pulled back to $\tilde Q$) on $C$.

At this point, by adding a constant to $B$, we could deform the
two-component curve $\Sigma'+C$ (which is singular where $\Sigma'$ and $C$ meet) back
to a smooth irreducible
curve of genus $g$ that does not pass through $P$ or $C$.  
Instead, we want to make the
transition to the Higgs branch. 

We recall that in the present paper, the curve $\Sigma$ is really an ingredient
in the description of a fivebrane in eleven dimensions.  The fivebrane propagates
in $\R^7\times \tilde Q$.  $\R^7$ has coordinates $x^0,x^1,\dots,x^3$ and $x^7,x^8,x^9$.
The fivebrane world-volume is of the form $\R^4\times \Sigma$, where
$\Sigma$ is a curve in $\tilde Q$ and $\R^4 $ is a subspace of $\R^7$ defined
by (for instance) $x^7=x^8=x^9=0$.

The transition to the Higgs branch can be described as follows.  When
$\Sigma$ degenerates to a curve that is a union of two branches $\Sigma'$ and $C$,
the fivebrane degenerates to two branches $\R^4\times \Sigma'$ and $\R^4\times C$.
At this point, it is possible for the two branches to move independently in $\R^7$.
$\R^4\times C$ can move to $\tilde {\R^4}\times C$, where
$\tilde {\R^4}$ is a {\it different} copy of $\R^4$ embedded in $\R^7$.  For unbroken
supersymmetry, $\tilde{\bf R^4}$ should be parallel to $\R^4$, so it is defined
in $\R^7$ by $(x^7,x^8,x^9)=\vec w$ for some constant $\vec w$.  

The four-dimensional field theory 
derived from a fivebrane on $\tilde{\bf R}^4\times C$
has no massless vector multiplets, as $C$ has genus zero.  It has one massless
hypermultiplet, whose components are $\vec w$ and $\int_C\beta$, where $\beta$ is
the chiral two-form on the fivebrane worldvolume.  

A motion of $\R^4\times \Sigma'$ in the $x^7,x^8,x^9$ directions, analogous
to the above, is {\it not} natural because $\Sigma'$ is non-compact and such
a motion would entail infinite action per unit volume on $\R^4$.
The  allowed motions of $\R^4\times \Sigma'$ are the motions of $\Sigma'$ in
$\tilde Q$ that determine the order parameters on the Coulomb branch
and that we have been studying throughout this paper.
The four-dimensional field theory derived from a fivebrane on $\R^4\times \Sigma'$
has $g-1$ massless 
vector multiplets, because $\Sigma'$ is a curve of genus $g-1$, and
one hypermultiplet.  The combined system of fivebranes on $\R^4\times \Sigma'$
and on $\tilde \R^4\times C$ 
has $g-1$ massless vector multiplets and one hypermultiplet.

There is no way to deform $\Sigma'$ to a curve of genus $g$.  It is only
$\Sigma'+C$ that can be so deformed.  So once $ C$
has moved to $\vec w=0$, there is no way to regain the $g^{th}$ massless vector
multiplet except by first moving $C$ back to $\vec w=0$.  The transition to the
Higgs branch has been made.

\subsec{Metric And Complex Structure}

Finally, using the techniques of \gibbons, we will briefly describe
how to exhibit the complex structure \juddo\ of the ALF manifold \turgo.
In that paper, the formula \turgo\ for the ALF hyper-Kahler
metric is obtained in the following way.

Let ${\bf H}$ be a copy of ${\bf R}^4$ with the flat hyper-Kahler metric.
Let ${\cal M}={\bf H}^d\times {\bf H}$, with coordinates $q_a$, $a=1,\dots, d$,
and $w$.  Consider the action on ${\cal M}$ of an abelian group $G$,
locally isomorphic to ${\bf R}^m$, for which the hyper-Kahler moment
map is
\eqn\ooplo{{\bf \mu}_a={1\over 2}{\bf r}_a+{\bf y},}
where ${\bf r}=q_ai\bar q_a$ and ${\bf y}=(w-\bar w)/2$.  Notation is as
explained in \gibbons.  $G$ is a product of $d$ factors; the $a^{th}$ factor,
for $a=1,\dots,d$, acts on $q_a$ by a one-parameter group of rotations that
preserve the hyper-Kahler metric, on $w$ by translations, and trivially on
the other variables.  The manifold  defined 
as ${\bf \mu}^{-1}({\bf e})/G$, with an arbitrary constant $\bf e$,
carries a natural hyper-Kahler metric, which is shown
in \gibbons\ to coincide with \turgo.  The choice of $\bf e$ determines
the positions $\vec{\bf x}_a$ of the sixbranes in \turgo.

To exhibit the structure of this hyper-Kahler manifold
as a complex manifold, one may proceed
as follows.  In any one of its complex structures, ${\bf H}$ can be identified
as ${\bf C}^2$.  One can pick coordinates so that each $q_a$ consists 
of a pair of complex variables $y_a,z_a$, and $w$ consists of a pair $v,v'$,
such that the action of $G$ is
\eqn\loopy{\eqalign{ y_a &  \to e^{i\theta_a}y_a \cr
                     z_a & \to e^{-i\theta_a}z_a\cr
                     v& \to v\cr
                     v'& \to v'-\sum_{a=1}^d\theta_a}}
where the $\theta_a$ are real parameters. 

Once a complex structure is picked, the moment map ${\bf \mu}$ breaks
up as a complex moment map $\mu_{\bf C}$ and a real moment map $\mu_{\bf R}$.
A convenient way to exhibit 
the complex structure of the ALF manifold is the following.  Instead of setting
${\bf \mu}$ to a constant value and dividing by $G$, one can set $\mu_\C$ to a constant
value and divide by $G_\C$, the complexification of $G$  (whose action is
given by the formulas \loopy\ with the $\theta_a$ now  complex-valued).\foot{
The quotient should be taken in the sense of geometric invariant theory.
This  leads to the fact, exploited in section 3.5,
 that when two sixbranes coincide in $v$ but
not in $x^6$, the ALF manifold \turgo\ is equivalent as a complex
manifold not to $yz=\prod_a(v-e_a)$ 
 but to the smooth resolution  $\tilde Q$ of that singular surface.
We will treat the invariant theory in a simplified way which misses the precise
behavior for $e_a=e_b$.  The calculation we do presently with invariants really
proves not that the ALF manifold is isomorphic to $\tilde Q$, but only that it 
has a holomorphic and generically one-to-one map to $\tilde Q$.
When $\tilde Q$ is smooth (as it is for generic $e_a$), 
the additional fact that the ALF manifold is hyper-Kahler implies that it must
coincide with $\tilde Q$. }      
The advantage of this procedure is that the complex structure is manifest.

The components of $\mu_{\bf C}$ are
\eqn\jiop{\mu_{\C,a}=y_az_a-v.}
Setting the $\mu_{\C,a}$ to constants, which we will call $-e_a$, 
means therefore taking
\eqn\iljop{y_az_a=v-e_a.}
Dividing by $G_\C$ is accomplished most simply by working with the
$G_\C$-invariant functions of $y_a,z_a,v$, and $v'$.  In other words, the
$G_\C$-invariants can be regarded as functions on the quotient $\tilde Q=
\mu_\C^{-1}(-e_a)/G_\C$. 

The basic invariants are $y=e^{iv'}\prod_{a=1}^dy_a$, $z=e^{-iv'}\prod_{a=1}^dz_a$,
and $v$.  The relation that they obey is, in view of \iljop,
\eqn\homely{yz=\prod_{a=1}^d(v-e_a),}
which is the formula by which we have defined the complex manifold $\tilde Q$.
This exhibits the complex structure of the ALF manifold, for generic sixbrane
positions $e_a$.

\newsec{Elliptic Models}

\subsec{Description Of The Models}

\midinsert
\centerline{\psfig{figure=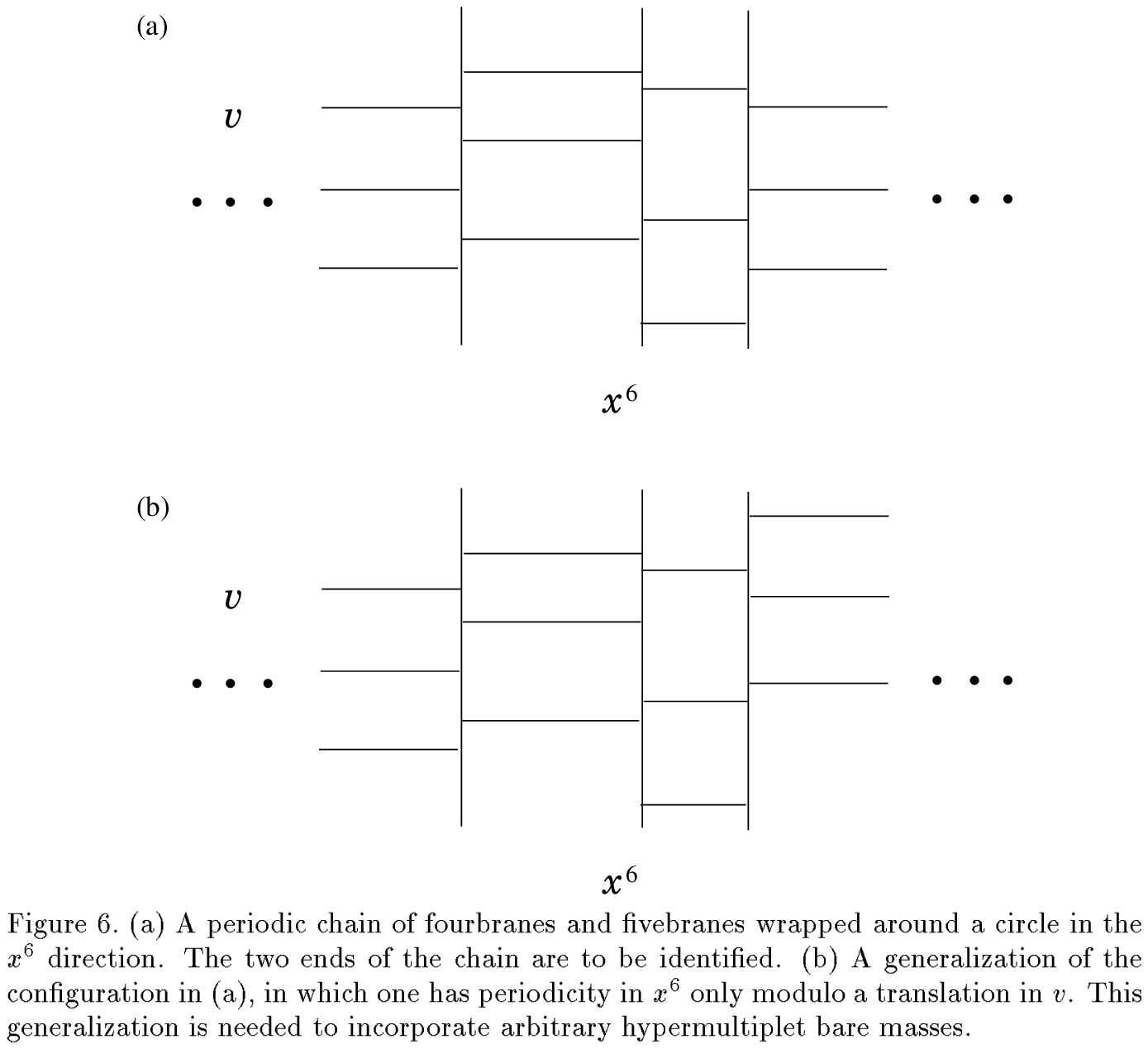,width=3.5in}}
\bigskip\bigskip\bigskip
\endinsert
In this section we compactify the $x^6$ direction to a circle, of radius $L$, 
and consider
a chain of $n$ fivebranes arranged around this circle, as in figure 6.
\foot{In the context of three-dimensional models with $N=4$ supersymmetry,
configurations of fivebranes arranged around a circle were studied in
\oret.}
Let $k_\alpha$ be the number of fourbranes stretching
between the $\alpha-1^{th}$ and $\alpha^{th}$
fivebrane, and let $d_\alpha$ be the number of sixbranes localized at points
between the $\alpha-1^{th}$ and $\alpha^{th}$ fivebrane.  The beta function of
the $SU(k_\alpha)$ factor in the gauge group is then
\eqn\indigo{b_{0,\alpha}=-2k_\alpha+k_{\alpha-1}+k_{\alpha+1}+d_\alpha.}
Since $\sum_\alpha b_{0,\alpha}=\sum_\alpha d_\alpha$, and the $d_\alpha$ are
all non-negative, the only case in which all beta functions are zero or
negative is that case that all $b_{0,\alpha}=d_\alpha=0$.
Then writing $0=\sum_\alpha k_\alpha(-2k_\alpha+k_{\alpha-1}+k_{\alpha+1})
=-\sum_\alpha(k_\alpha-k_{\alpha-1})^2$, we see that
this occurs if and only if  all $k_\alpha$ are equal to a fixed integer
$k$.  The present section will be devoted to analyzing this case.

The gauge group is   $G=U(1)\times SU(k)^n$.  Only the occurrence of
a $U(1)$ factor requires special comment.  The condition \koggo\ ``freezes 
out''
the difference between the $U(1)$ factors in the gauge group supported on 
alternate
sides of any given fivebrane.  In sections 2 and 3, we considered a finite 
chain
of fivebranes with $U(1)$'s potentially supported only in the ``interior'' of 
the chain,
 and this condition sufficed to eliminate
all $U(1)$'s.  In the present case of $n$ fivebranes arrranged around a circle
with fourbranes connecting each neighboring pair, \koggo\ eliminates $n-1$ of 
the
$U(1)$'s, leaving a single (diagonal) $U(1)$ factor in the gauge group.

\def\1{{\bf 1}}
Hypermultiplets arise  from fourbranes that meet a single fivebrane
at the same point in space from opposite sides.  If the symbol $\k_{\alpha}$
represents the fundamental representation of the $\alpha^{th}$ $SU(k)$ factor
in $G$, then the hypermultiplets transform as
$\oplus_{\alpha=1}^n\k_\alpha\otimes 
\bar\k_{\alpha+1}$.
Note that all of these hypermultiplets are neutral under the $U(1)$, so that
all beta functions vanish including that of the $U(1)$.  The $U(1)$, while 
present,  is thus completely decoupled in the model.   The
curve $\Sigma$ that we will eventually construct will have the property that
its Jacobian determines the coupling constant of the $U(1)$ factor as well as
the structure of the $SU(k)^n$ Coulomb branch.

A special case that merits some special discussion is the case $n=1$.  In that
case the gauge group consists just of a single $SU(k)$ (times the decoupled 
$U(1)$)
and the $\k\otimes \bar \k$ hypermultiplet consists of a copy of the adjoint
representation of $SU(k)$ plus a neutral singlet.  This in fact corresponds to 
the $\N=4$
theory with gauge group $U(k)$; however, we will study it eventually in the 
presence
of a hypermultiplet bare mass that breaks $\N=4$ to $\N=2$. Precisely this
model has been solved
in \donagiwitten, and we will recover the 
description in  that paper.

\bigskip\noindent{\it Hypermultiplet Bare Masses}

Before turning to $M$ theory, 
we will analyze, in terms of Type IIA, the hypermultiplet bare masses.

Let $a_{i,\alpha}$, $i=1,\dots,k$ be the $v$ values of the fourbranes between
the $\alpha-1^{th}$ and $\alpha^{th}$ fivebranes.  According to \jumper, the 
bare
mass $m_\alpha$ of the $\k_\alpha\otimes \bar\k_{\alpha+1}$ hypermultiplet
is 
\eqn\hurgo{m_\alpha={1\over k}\left(\sum_ia_{i,\alpha}-\sum_ja_{j,\alpha+1}\right).}

This formula seems to imply that the $m_\alpha$ are not all independent, but
are restricted by $\sum_\alpha m_\alpha=0$.  However, that restriction can be
avoided if one choses correctly the spacetime in which the branes propagate.

So far, we have described  the positions of the fourbranes and fivebranes
in terms of $x^6$ and $v=x^4+ix^5$.  Since we are now compactifying the $x^6$
direction to a circle, this part of the spacetime is so far ${ T}=\S^1\times 
\C$,
where $\S^1$ is the circle parametrized by $x^6$ and $\C$ is the $v$ plane.

We can however replace $\S^1\times \C$ by a certain $\C$ bundle over $\S^1$.
In other words, we begin with $x^6$ and $v$ regarded as coordinates on
$\R^3=\R\times \C$, and instead of dividing simply by $x^6\to x^6+2\pi L$ for
some $L$, we divide by the combined operation
\eqn\joko{\eqalign{x^6 & \to x^6+ 2\pi L \cr
                   v & \to v + m ,\cr}}
for an arbitrary complex constant $m$. 
Starting with the flat metric on $\R^3$,
this gives a $\C$ bundle over $\S^1$ with a flat metric; we call this space 
$T_m$.
 Now when one goes all the way around
the $x^6$ circle, one comes back with a shifted value of $v$, as suggested in
figure 6(b). The result is
that the formula $\sum_\alpha m_\alpha=0$ which  one would get on $\R\times 
\C$
is replaced on $T_m$ by
\eqn\koko{\sum_\alpha m_\alpha = m.}
Thus arbitrary hypermultiplet bare masses are  possible, with a judicious 
choice of
the spacetime.

\subsec{Interpretation In $M$ Theory}

Now we want to study these models via $M$ theory.

Going to $M$ theory means first of all including another circle, parametrized
by a variable $x^{10}$ with $x^{10}\cong x^{10}+2\pi R$.
Now because in the present section we are compactifying also the $x^6$ 
direction
to a circle,   we have really two circles.  The metric structure, however,
need not be a simple product $\S^1\times \S^1$.  Dividing $x^6\to x^6+2\pi L$ 
can be accompanied by a shift of $x^{10}$, the combined operation being
\eqn\kopo{\eqalign{x^6& \to x^6+2\pi L\cr
x^{10} & \to x^{10}+\theta R\cr}}
with some angle $\theta$.  
We also still divide by $x^{10}\to x^{10}+2\pi R$, as in uncompactified Type 
IIA.
In the familiar complex structure in which $s=x^6+ix^{10}$ is holomorphic,
the quotient of the $s$ plane by these equivalences is a complex Riemann 
surface
$E$ of genus one which -- by varying $L$ and $\theta$ for fixed $R$ (that is 
fixed ten-dimensional
Type IIA string coupling constant) -- can have an arbitrary complex structure.
$E$ also has a flat metric with an area that (if we let $R$ vary) is arbitrary;
this, however, will be less important, since we are mainly studying properties
that are controlled by the holomorphic data.

The interpretation of this generalization for our problem of gauge theory on
branes is as follows.  The $\alpha^{th}$ fivebrane has, 
in the $M$ theory description,
a position $x^{10}_\alpha $ in the $x^{10}$ direction, as 
well as a position $x^6_{\alpha}$
in the $x^6$ direction.  The theta angle $\theta_\alpha$ of the
$\alpha^{th}$  $SU(k)$ factor in the gauge group is 
\eqn\oook{\theta_\alpha = {x^{10}_{\alpha}
-x^{10}_{\alpha-1}\over R}.}   If metrically $x^6-x^{10}$ space 
were a product $\S^1\times \S^1$
(or in other words if $\theta=0$ in \kopo) then   \oook\ would imply that
$\sum_\alpha\theta_\alpha=0$.  Instead, via \kopo, we arrange that 
when one goes around a circle in the $x^6$ direction, one comes back with a 
shifted
valued of $x^{10}$; as a result  one has
\eqn\jggo{\sum_\alpha\theta_\alpha=\theta.}

In a Type IIA description, one would not see the $x^{10}$ coordinate.
The fact that $x^{10}$ shifts by $\theta$  under $x^6\to x^6+2\pi L$ would
be expressed by saying that the holonomy around the $x^6$ circle of the 
Ramond-Ramond
$U(1)$ gauge field of Type IIA is $e^{i\theta}$.  The $x^{10}$ positions of
a fivebrane would be coded in the value of a certain scalar field that
propagates on the fivebrane.

\bigskip\noindent{\it Duality Group}

In general, $E$ is a (smooth) genus one Riemann surface with an arbitrary 
complex
structure, and the fivebranes are at $n$ arbitrary points $p_1,\dots, p_n$ on 
$E$.
By varying in an arbitrary fashion the complex structure of $E$ and the choice
of the $p_\sigma$, the bare couplings and theta angles of 
$G'=\prod_{\alpha=1}^k SU(k)$
can be varied in  arbitrarily.  (The  coupling and theta angle of the 
$U(1)$ 
factor in the full gauge group $G=U(1)\times G'$ is then determined 
in terms of those.)  The duality group of these models can thus be described
as follows.  Let ${\cal M}_{1,n}$ be the moduli space of smooth Riemann 
surfaces
of genus one with $n$ distinct, unordered marked points.  The duality
group is then $\pi_1({\cal M}_{1,n})$.  For $n=1$, $\pi_1({\cal M}_{1,1})$ is
the same as $SL(2,\Z)$, and this becomes the usual duality group
of $\N=4$ super Yang-Mills theory.  For $n>1$, $\pi_1({\cal M}_{1,n})$ is
a sort of hybrid of $SL(2,\Z)$ and the duality group found in section 3.

\bigskip\noindent{\it Incorporation Of $v$}

We now want to consider also the position of the fivebranes in $v=x^4+ix^5$.
An important special case is that in which the fivebranes propagate in
$X=E\times \C$, where $\C$ is the complex $v$ plane.  However, from the 
discussion
of \joko, it is clear that in general we should consider not a product $E\times
\C$ but a $\C$ bundle over $E$.  In general, we start with $\R\times 
\S^1\times\C$
(with respective coordinates $x^6$, $x^{10}$, and $v$) and divide by the
combined symmetry
\eqn\ruby{\eqalign{x^6    &    \to x^6+2\pi L \cr
                   x^{10} &    \to x^{10} +\theta \cr
                   v      &    \to      v+ m.\cr}}
The quotient is a complex manifold that we will call $X_m$; it can be regarded
as a $\C$ bundle over $E$.  From the discussion at the Type IIA level, it is 
clear
that the parameter $m$ must be identified with the sum of the hypermultiplet
bare masses.

The complex manifold $X_m$ will actually not enter as an abstract complex
manifold\foot{As such it is isomorphic to $\C^*\times \C^*$.}; 
the map $X_m\to E$ (by forgetting $\C$) will be an important part of the 
structure.  As a $\C$ bundle over $E$, $X_m$ is an ``affine bundle''; this
means that the fibers are all copies of $\C$ but there is no way to globally
define an ``origin'' in $\C$, in a fashion that varies holomorphically.
Such affine bundles over $E$, with the associated complex line bundle (in which
one ignores shifts of the fibers) being trivial, are classified by 
the sheaf cohomology group $H^1(E,{\cal O}_E)$, which is one-dimensional;
the one complex parameter that enters is what we have called $m$.  If
 $X_m$ is viewed just as a complex manifold with map to $E$, $m$ could be set 
to 1
(given that it is non-zero) by rescaling $v$, but we prefer not to do that
since the fivebrane effective action is not invariant under rescaling of $v$.

The complex manifold $X_m$ appeared  in \donagiwitten, where the $SU(k)$ theory
with massive adjoint hypermultiplet -- in other words, the
$n=1$ case of the series of models considered here -- was described in
terms of an appropriate curve in $X_m$, rather as we will do below.
Actually, in what follows we will consider  curves in $X_m$ that ``go to 
infinity''
at certain points, corresponding to the positions of fivebranes.  In
\donagiwitten, a ``twist'' of $X_m$ was made to keep the curve from going to
infinity.

\subsec{Solution Of The Models} 

What remains is to describe the solution of the models.
First we consider the special case that
the sum of the hypermultiplet bare masses is zero,
\eqn\ilpo{\sum_\alpha m_\alpha = 0,}
so that the model will be described by a curve $\Sigma$ in $X=E\times \C$.
There are $n$ fivebranes at points $p_1,p_2,\dots,p_n$ in $E$; and to
use a classical Type IIA language (which we will presently reformulate in a way
more suitable in $M$ theory) each pair
of adjacent five-branes is connected by $k$ fourbranes.

First of all, the elliptic curve $E$ can be described by a Weierstrass 
equation,
$zy^2=4x^3-g_2xz^2-g_3z^3$ in homogeneous coordinates $x,y,z$; $g_2$ and $g_3$
are complex constants.  Usually we work in coordinates with $z=1$ and
write simply
\eqn\prlim{y^2=4x^3-g_2x-g_3.}
$E$  admits an everywhere non-zero
holomorphic differential
\eqn\kko{\omega={dx\over y}.}

To incorporate the classical idea that there are $k$ fourbranes between each
pair of fivebranes, we proceed as follows.
$X$ maps to $E$ by forgetting $\C$; under this map, the curve $\Sigma\subset X$
maps to $E$.
Via the map $\Sigma\to E$, $\Sigma$ can be interpreted as a $k$-fold cover
of $E$, the $k$ branches being the positions of the fourbranes in $\C$.  In
other words, $\Sigma$ is defined by an equation $F(x,y,v)=0$, where $F$ is
of degree $k$ in $v$:
\eqn\uggu{F(x,y,v)=v^k-f_1(x,y)v^{k-1}+f_2(x,y)v^{n-2}\mp \dots 
+(-1)^kf_k(x,y).}  
The functions $f_i(x,y)$
are meromorphic functions on $E$ (and hence are rational functions of $x$ and 
$y$)
obeying certain additional conditions that will be described.

The idea here  is that for generic $x$ and $y$, 
the equation $F(x,y,v)$ has $k$ roots for $v$, which are the positions of
the fourbranes in the $v$ plane.  Call those roots $v_i(x,y)$.
Unless the $f_i$ are all constants, there will be
points on $E$ at which some of the $f_i$ have poles.  At such a point, at least
one of the $v_i(x,y)$ diverges.

We would like to interpret the poles in terms of positions of fivebranes.
Let us first explain why such an interpretation exists.   
 An $M$ theory fivebrane located at $v=v_0$ would be 
interpreted in Type IIA as a  fourbrane at $v=v_0$.  A Type IIA fivebrane
 located at some point $p\in E$ also corresponds to a fivebrane in 
Type IIA.
The equation for such a fivebrane is, say, $s=s_0$ where $s$ is a local 
coordinate
on $E$ near $p$ and $s=s_0$ at $p$.  The combined Type IIA fourbrane-fivebrane
system can be described in $M$ theory by a fivebrane with the world-volume
\eqn\iloop{(v-v_0)(s-s_0)=0.}
The space of solutions of this equation
has two branches, $v=v_0$ and $s=s_0$; these are
interpreted in Type IIA as the fourbrane and fivebrane, respectively.  There
is a singularity where the two branches meet.
Now without changing the asymptotic behavior of the curve described in \iloop\
-- in fact, while changing only the microscopic details -- one could add a 
constant to the equation, getting
\eqn\jiloop{(v-v_0)(s-s_0) =\epsilon.}
The singularity has disappeared; what in Type IIA is a fourbrane and a 
fivebrane appears in this description as a single, smooth, irreducible object.
On the other hand, if we solve \jiloop\ for $v$ we get
\eqn\hiloop{v=v_0+{\epsilon\over s-s_0}.}
We see that a fivebrane corresponds to a simple (first order) pole in $v$.

Poles of the $f_i$ will lead to singularities of the $v_i$.
It is now possible to determine what kind of singularities we should allow
in the $f_i$.  At a point $p_\sigma$ at which a fivebrane is located, one
of the $v_i$ should have a simple pole, analogous to that in \hiloop, and the
others should be regular.
The $v_i$ will behave in this way if and only if the $f_i$ have simple
poles at $p_\sigma$.
So  {\it the functions $f_1,\dots, f_k$ have simple
poles at the points $p_1,\dots,p_n$ and no other singularities}.

This then almost completes the description of the solution of the models: they
are described by curves $F(x,y,v)=0$ in $E\times \C$, where $F$ is as in \uggu\
and the allowed functions $f_i$ are characterized by the property just stated.
What remains is to determine which parameters in the $f_i$ are hypermultiplet
bare masses and which ones are order parameters describing the choice of a 
quantum
vacuum. 

First let us count all parameters.  By the Riemann-Roch theorem,
the space of meromorphic functions on
$E$ with simple poles allowed at $p_1,\dots, p_n$ is $n$-dimensional.  As
we have $k$ such functions, there are $kn$ parameters in all.  Of these,
$n-1$ should be hypermultiplet bare masses (because of \ilpo\ there are only
$n-1$ hypermultiplet bare masses),
leaving $n(k-1)+1$ order parameters.  The gauge group $G=U(1)\times SU(k)^n$
has rank $n(k-1)+1$, so $n(k-1)+1$ is the dimension of the Coulomb branch,
and hence is the correct number of order parameters.
It remains then to determine which $n-1$ parameters are the hypermultiplet
bare masses.

Let us note the following interpretation of the function $f_1$:
in view of the factorization $F(x,y,v)=\prod_{i=1}^k(v-v_i(x,y))$, one has
\eqn\olopo{f_1(x,y)=\sum_{i=1}^k v_i(x,y).}
The generic behavior is that near any one of the $p_\sigma$, all
of the $v_i$ except one remain finite, and the remaining one, say $v_1(x,y)$,
has a simple pole.  
So according to \olopo\ the singular behavior of $v_1$ is the same as the
singular behavior of $f_1$.  In other words, the singular part of $f_1$ 
determines
the behavior of $\Sigma$ near infinity.
Since hypermultiplet bare masses are always
coded in the behavior of the curve $\Sigma$ at infinity -- as we saw in \boggo,
that is why the bare masses are constant -- the hypermultiplet bare
masses must be coded in the singular part of $f_1$.  

The singular part of $f_1$ depends only on $n-1$ complex parameters.  
In fact, $f_1$ itself depends on $n$ complex parameters, but as one is free
to add a constant to $f_1$ without affecting its singular behavior, the 
singular
part of $f_1$ depends on $n-1$ parameters.  Thus, fixing the hypermultiplet
bare masses completely fixes the singular part of $f_1$.  The additive constant
in $f_1$ and the parameters in $f_j$, $j>1$ are the order parameters specifying
a choice of quantum vacuum.  Actually, the additive
constant in $f_1$ is the order parameter on the Coulomb branch of the $U(1)$
factor in the gauge group; 
this constant can be shifted  by adding
a constant to $v$ and so does not affect the Jacobian of $\Sigma$, in agreement
with the fact that the $U(1)$ is decoupled.  The order parameters of
the $SU(k)^n$ theory are the $n(k-1)$ coefficients in $f_2,f_3,\dots, f_n$.

To be more complete, one would like to know which functions of the
singular part of $f_1$ are the hypermultiplet bare masses $m_\alpha$.  
One approach to this question is 
to think about the integrable system that controls the  structure of the 
Coulomb
branch.  We recall from section 2.3 that a point in the phase space of
this integrable system is given by the choice of a curve $\Sigma\subset E\times\C$
 with fixed
behavior at infinity together with the 
choice of a line bundle on the compactification
of $\Sigma$.
As in section 17 of 
the second paper in \swtwo,
 the cohomology
class of the complex symplectic form on the phase space should vary
linearly with the masses.  How to implement this
condition for integrable systems of the kind considered here is explained
in section 2 of \donagiwitten.  The result is as follows:
the hypermultiplet bare masses are the residues of the differential
form $\beta = f_1(x,y)\omega$.  Since the sum of the residues of a meromorphic
differential form vanishes, this claim is
in accord with \ilpo.  

\subsec{Extension To Arbitrary Masses}

What remains is to eliminate the restriction \ilpo\ and solve the
models with arbitrary hypermultiplet bare masses.  For this, as we have
discussed in section 4.2, it is necessary to consider curves $\Sigma$ not in 
$X=E\times \C$, but in an affine bundle over $E$ that we have called $X_m$.

$X_m$ differs from the trivial product bundle $X= 
E\times \C\to E$ by twisting by 
an element
of $H^1(E,{\cal O}_E)$.  That cohomology group vanishes if a point is
deleted from $E$.  We can pick that point to be the point $p_\infty$ with
 $x=y=\infty$  in the
Weierstrass model \prlim.  To preserve the symmetry among the points
$p_\sigma$ at which there are fivebranes, we take $p_\infty$ to be
distinct from all of the $p_\sigma$. 
Because $X_m$ coincides with $X$ away from the fiber over $p_\infty$,
we can describe the curve $\Sigma$ away from 
$p_\infty$
by the same equation as before, $F(x,y,v)=0$
with
\eqn\jomo{F(x,y,v)=v^k-f_1(x,y)v^{k-1} +f_2(x,y)v^{k-2}\mp \dots 
+(-1)^kf_k(x,y).}
Away from $x=y=\infty$, 
the functions $f_i(x,y)$ are subject to the same conditions as before
-- no singularities except simple poles at the points $p_\sigma$.

Previously, we required that the roots $v_i(x,y)$ were finite at $x=y=\infty$
(since there are no fivebranes there) and hence that the $f_i$ were finite
at $x=y=\infty$.  For describing a curve on $X_m$, that is not the right
condition.  The trivialization of the affine bundle $X_m$ over $E$ minus
the point at infinity breaks down at $x=y=\infty$.  A good coordinate
near infinity is not $v$ but 
\eqn\tomo{\tilde v = v +\left({m\over 2k}\right){y\over x}.}
(Instead of $y/x$ one could use any other function with a simple pole
at $x=y=\infty$. For the moment one should think of the $m/2k$ on the right
hand side \tomo\ as an arbitrary constant.)  
It is not $v$ but $\tilde v$ that should be finite
at $x=y=\infty$.  

Thus the restrictions on the $f_i$ that are needed to solve the model with
arbitrary hypermultiplet bare masses can 
be stated as follows:

(1) The functions $f_i(x,y)$ are meromorphic functions on $E$ with no 
singularities
except simple poles at the $p_\sigma$, $\sigma=1,\dots, n$, and poles
(of order $i$ ) at $x=y=\infty$.

(2) The singular part of the function $F(v,x,y)$ near $x=y=\infty$
disappears if this function is expressed in terms of $\tilde v$ instead of $v$.

The hypermultiplet bare masses $m_\alpha$ are the residues of the differential
form $\beta=f_1 \omega$ at the points $p_\sigma$.  Since the sum of the 
residues
of $\beta$ will vanish, $\beta$ has a pole at $x=y=\infty$ with residue
$-\sum_\alpha m_\alpha$.  We can now relate this expression to the parameter
$m$ in \tomo.  Since condition (2) above implies that the singular
behavior of $f_1$ is $f_1=-my/2x+\dots, $ and since the 
differential form $(dx/y)(y/2x)$ has a pole at infinity with residue 1, the 
residue
of $\beta $ is  in fact $-m$, so we get
\eqn\jogoc{m=\sum_\alpha m_\alpha.}
This relation between the coefficient $m$ by which $X_m$ is twisted and the
hypermultiplet bare masses $m_\alpha$ was anticipated in \koko.

Just as in the case $m=0$ that we considered first, the order parameters
on the Coulomb branch are the parameters not fixed by specifying the singular
part of $f_1$.

In \donagiwitten, the solution of this model for the
special case $n=1$ was expressed in an equivalent but
slightly different way.  Since -- to adapt the discussion to the present
language -- there was only one fivebrane, the fivebrane was placed at 
$p_\infty$  without any loss of symmetry.  In place of conditions
 (1) and (2), the requirements on the $f_i$ were 
the following:

($1'$) The functions $f_i(x,y)$ are meromorphic functions on $E$ with no 
singularities
except a pole of order at most $i$ 
 at $x=y=\infty$.

($2'$) After the change of variables \tomo, the singularity of the function
$F(x,y,v)$ at $x=y=\infty$  is only a simple pole.

These conditions were used as the starting point for fairly detailed 
calculations of the properties of the model.

For the general case of $n$ fivebranes,
if we choose  one of the fivebrane locations, say   $p_1$, 
to equal $p_\infty$, then (1) and (2) can be replaced by the following 
conditions:

($1''$) The functions $f_i(x,y)$ are meromorphic functions on $E$ 
whose possible singularities are simple poles at $p_2,\dots,p_n$ and a pole
of order $i$ at $x=y=\infty$.

($2''$) After the change of variables \tomo, the singularity of the
function $F(x,y,t)$ at $x=y=\infty$ is only a simple pole.

These conditions are equivalent to (1) and (2), up to a translation on
$E$ that moves $p_1$ to infinity and a change of variables $v\to v+a(x,y)$
for some function $a$.
\listrefs
\end